\documentclass[journal,twoside]{IEEEtran}
\usepackage{setspace}
\usepackage{balance}
\usepackage{amsmath}
\usepackage{bm}
\usepackage{amssymb}
\usepackage{gensymb}
\usepackage{siunitx}
\usepackage{cite}
\usepackage{mathtools}
\usepackage[super]{nth}
\usepackage{bbm}
\usepackage{graphicx}
\usepackage{xcolor}
\usepackage{epstopdf}
\usepackage{xfrac}
\usepackage{nicefrac}
\usepackage{dblfloatfix}
\usepackage{array}
\usepackage{float}
\usepackage[export]{adjustbox}
\usepackage{lipsum}
\usepackage{tikz}
\usepackage[T1]{fontenc}
\usepackage{multirow}
\usepackage{multicol}
\usepackage{nicematrix}
\usepackage{hyperref}
\usepackage{times}

\usepackage{comment}

\usepackage{xpatch}
\makeatletter
\chardef\TPT@@@asteriskcatcode=\catcode`*
\catcode`*=11
\xpatchcmd{\threeparttable}
  {\TPT@hookin{tabular}}
  {\TPT@hookin{tabular}\TPT@hookin{tabu}}
  {}{}
\catcode`*=\TPT@@@asteriskcatcode
\makeatother

\ifCLASSOPTIONcompsoc
\usepackage[caption=false,font=normalsize,labelfont=sf,textfont=sf]{subfig}
\else
\usepackage[caption=false,font=footnotesize]{subfig}
\fi

\begin{document}
\title{Generalized Gain and Impedance Expressions for Single-Transistor Amplifiers}

\author{Brian Hong\textsuperscript{\href{https://orcid.org/0000-0001-8099-0312}{\includegraphics[width=2.25ex]{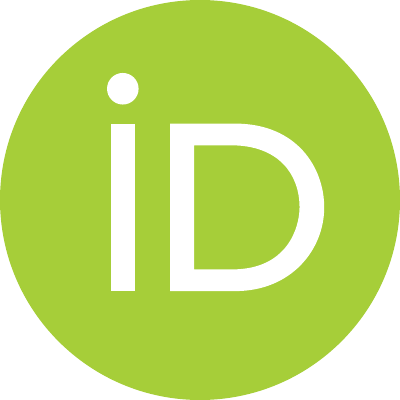}}},~\IEEEmembership{Member,~IEEE}
\thanks{The author was with the Department of Electrical Engineering, California Institute of Technology, Pasadena, CA 91125 USA, and with Yale Law School, Yale University, New Haven, CT 06511 USA. He is now with Broadcom Inc., San Jose, CA 95134 USA (e-mail: brian.hong@broadcom.com).}}

\markboth{}{Hong: Generalized Expressions for Single-Transistor Amplifiers}

\maketitle

\begin{abstract}
This expository manuscript presents generalized expressions for the low-frequency voltage gain and terminal impedances of each of the three fundamental bipolar-amplifier topologies (i.e., common emitter, common base, and common collector). Unlike the formulas that students typically learn and designers typically use, the equations presented in this tutorial assume the most general set of conditions: finite output resistance and base-collector current gain, a load resistor at each non-input terminal of the transistor, and a ``feedback'' resistor between the base and collector terminals. Although perhaps algebraically complex at first glance, emphasis is placed on mathematical elegance and ease of use---expressions are formulated in terms of sub-terms that capture important aspects of the circuit's behavior. Similarities in the mathematical structure of the results reveal a deeper conceptual connection between different amplifier topologies and, ultimately, a reciprocity relationship between the base and emitter terminals. Familiar approximate expressions are subsumed as special cases. Tables consolidating the expressions in an organized fashion are provided. Companion results for metal-oxide-semiconductor (MOS) single-transistor amplifiers are also included.
\end{abstract}


\IEEEpeerreviewmaketitle

\section{Introduction}
\IEEEPARstart{T}{ransistors} are the building blocks of modern technology. An understanding of their amplification abilities lies at the heart of the vast majority of microelectronics curricula \cite{Razavi:fund,Razavi:CMOS,GrayMeyer,SedraSmith,Searle,NetwAnal,Ali} and serves as the intellectual backbone of a career in chip design and silicon development. However, for students, the subject of analog circuits may seem chock full of equations and approximations---difficult to remember and even more difficult to use. Although seasoned designers learn to rely on intuition---familiarity with certain topologies enables prediction of their behavior---this paper prescribes another antidote to the chaos. Not only do we provide a uniform generalization of the gain and impedance relationships that are typically taught to students, but our clean, comprehensive, and organized compilation of those results also allows underlying similarities in mathematical form (and therefore parallels in physical behavior) to be readily observed.

Specifically, this tutorial paper will go through the low-frequency voltage gains of the common emitter, common base, and common collector amplifiers, as well as the low-frequency impedances looking into each of the transistor's three terminals.\footnote{The impedance ``looking into'' a terminal is defined as the impedance between that terminal and the ground terminal (in our case, small-signal ground). Also, we will use the terms ``impedance'' and ``resistance'' interchangeably since they are identical in low-frequency settings.} For each analysis, we start with the ``most general'' case---a resistor loads each (non-input) terminal, and a ``feedback'' resistance connects the base and collector terminals---before running through a set of simplifying assumptions. Along the way, we will uncover and explore a ``reciprocity'' relationship---a symmetry of sorts that manifests itself in different ways in different parts of our discussion. Finally, the corresponding expressions for the MOS-transistor analogues of these circuits (with the body effect included\footnote{The body effect is accounted for in the form of the small-signal back-gate transconductance $g_{mb}$ \cite{Razavi:CMOS,GrayMeyer,SedraSmith,Ali}. Obviously, if the body effect is neglected (i.e., $g_{mb}=0$), the MOS expression can be obtained by simply taking \hbox{$\beta\rightarrow\infty$} (i.e., $\alpha=1$) in the bipolar expression.}) are provided, and simulation results for a 36-nm $n$-channel FinFET are shown.

One might argue that expressions at this level of accuracy are not worth their complexity---and are therefore unlikely to be used by designers, especially as computing resources become more abundant and circuit simulators become more sophisticated. Moreover, linearized small-signal analyses possess limited modeling power to begin with, an increasingly apparent phenomenon as processes continue to scale down. Still, from an intellectual standpoint, these expressions help form a complete understanding of how a single transistor transforms small-signal voltages and currents across its three terminals. In doing so, these equations elucidate the conditions under which certain design choices (e.g., degeneration, feedback) or non-idealities (e.g., channel-length modulation) dominate the circuit's behavior. Our analysis therefore weaves together the seemingly disparate approximations that hold when these effects are individually or jointly ignored.

A couple preliminaries and notational matters. Throughout this paper, the ``$\parallel$'' symbol denotes parallel combination:
\begin{equation*}
R_A \parallel R_B \coloneqq \dfrac{1}{\dfrac{1}{R_A} + \dfrac{1}{R_B}} = \dfrac{R_AR_B}{R_A+R_B}.
\end{equation*}
Also, $r_m \coloneqq 1/g_m$ and $r_{mb} \coloneqq 1/g_{mb}$ are defined as the resistances whose values are equal to the inverses of the transconductance $g_m$ and the back-gate transconductance $g_{mb}$, respectively.

\section{Common Emitter Amplifier}
Consider the common emitter amplifier whose schematic is shown in the first row of Table~\ref{tab:CE}. The input $v_{\mathrm{in},b}$ is applied at the base, and the output $v_c$ is taken from the collector. Using the transistor's $\pi$-model, we can write down the nodal equations to be
\begin{equation}
\begin{aligned}
\begin{pmatrix}
\dfrac{1}{R_C} + \dfrac{1}{R_F} + \dfrac{1}{r_o} & -g_m - \dfrac{1}{r_o} \\[1em]
-\dfrac{1}{r_o} & g_m + \dfrac{1}{r_{\pi}} + \dfrac{1}{R_E} + \dfrac{1}{r_o}
\end{pmatrix}
\begin{pmatrix}
v_c \\[0.5em] v_e
\end{pmatrix}
\\[1em] =
\begin{pmatrix}
-g_m + \dfrac{1}{R_F} \\[1em]
g_m + \dfrac{1}{r_{\pi}}
\end{pmatrix}
v_{\mathrm{in},b}.
\end{aligned}
\label{eq:CE-nodal}
\end{equation}
Solving this matrix equation \eqref{eq:CE-nodal} for the gain, $A_v^{\mathrm{CE}} \coloneqq v_c/v_{\mathrm{in},b}$, yields the general result in the first row of Table~\ref{tab:CE}. This formula, although perhaps formidable in its appearance, actually admits a highly intuitive interpretation. The term $1 + g_mR_E/\alpha + R_E/r_o$ can be viewed as an emitter-degeneration factor that degrades the transconductance $g_m$ but ``enhances'' the output resistance $r_o$. The feedback resistor $R_F$ serves the usual role of decreasing the (effective) transconductance by $1/R_F$ and appearing in parallel with the load. That is,
\begin{equation}
A_v^{\mathrm{CE}} = - \left(g_{m,\mathrm{eff}}^{\mathrm{CE}} - 1/R_F\right) \left(R_C \parallel R_F \parallel r_{o,\mathrm{eff}}^{\mathrm{CE}}\right),
\label{eq:CE_intuitive}
\end{equation}
where
\begin{equation*}
g_{m,\mathrm{eff}}^{\mathrm{CE}} = g_m \left(\dfrac{1-\dfrac{R_E}{\beta r_o}}{1+\dfrac{g_mR_E}{\alpha}+\dfrac{R_E}{r_o}}\right)
\end{equation*}
and
\begin{equation*}
r_{o,\mathrm{eff}}^{\mathrm{CE}} = r_o \left(\dfrac{1+\dfrac{g_mR_E}{\alpha}+\dfrac{R_E}{r_o}}{1+\dfrac{g_m R_E}{\beta}}\right).
\end{equation*}
The next three rows provide various approximations of interest: no degeneration, no feedback, and no output resistance. The gain of the MOS counterpart to the common emitter amplifier---the common source amplifier---is given in the first row of Table~\ref{tab:MOS}.

\begin{table*}[ht]
\centering
\caption{Common Emitter Amplifier -- Voltage Gain}
\label{tab:CE}
    \begin{NiceTabular}[width=\linewidth]{X[13,c,m] X[15,c,m] X[33,c]}[hvlines]
    General Expression & \vspace{2pt} \includegraphics[scale=0.4]{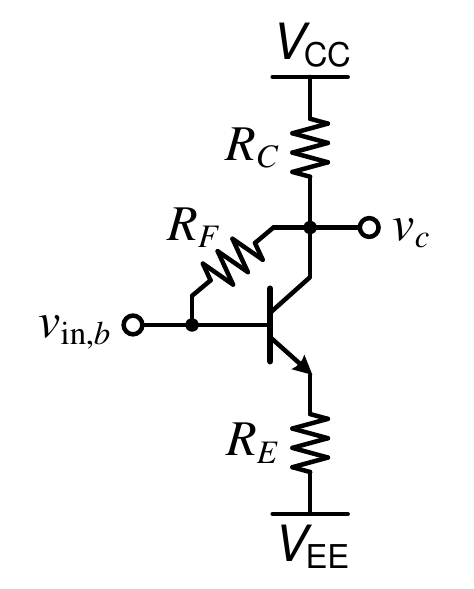} & \(\displaystyle A_v^{\mathrm{CE}} = -\dfrac{g_m\left(\dfrac{1-\dfrac{R_E}{\beta r_o}}{1+\dfrac{g_mR_E}{\alpha}+\dfrac{R_E}{r_o}}\right)-\dfrac{1}{R_F}}{\dfrac{1}{R_C \parallel R_F} + \dfrac{1}{r_o}\left(\dfrac{1+\dfrac{g_m R_E}{\beta}}{1+\dfrac{g_mR_E}{\alpha}+\dfrac{R_E}{r_o}}\right)} \) \\
    No Degeneration \linebreak $\left(R_E=0\right)$ & \vspace{2pt} \includegraphics[scale=0.4]{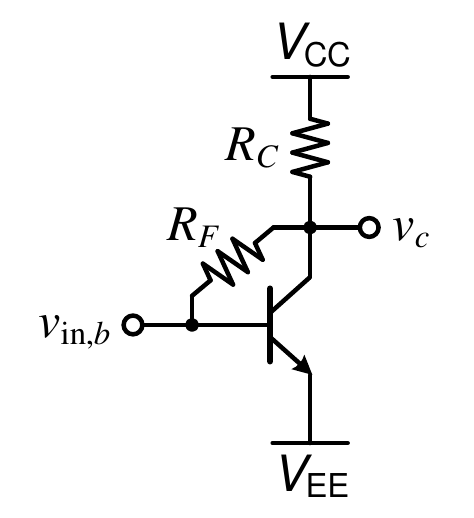} & \(\displaystyle A_v^{\mathrm{CE}} = -\left(g_m-\dfrac{1}{R_F}\right) \left(R_C \parallel R_F \parallel r_o\right) \) \\
    No Feedback \linebreak $\left(R_F \rightarrow \infty\right)$ & \vspace{2pt} \includegraphics[scale=0.4]{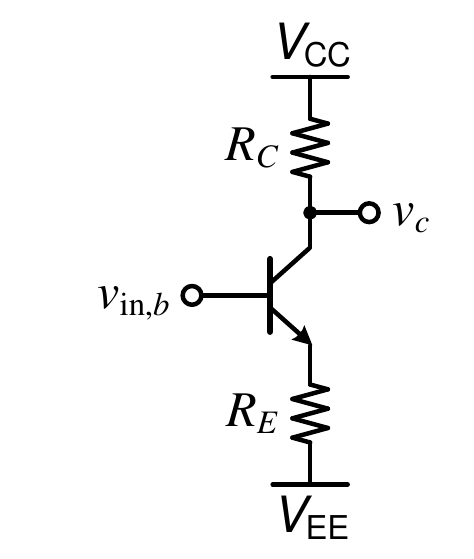} & \(\displaystyle A_v^{\mathrm{CE}} = -\dfrac{g_m\left(\dfrac{1-\dfrac{R_E}{\beta r_o}}{1+\dfrac{g_mR_E}{\alpha}+\dfrac{R_E}{r_o}}\right)}{\dfrac{1}{R_C} + \dfrac{1}{r_o}\left(\dfrac{1+\dfrac{g_m R_E}{\beta}}{1+\dfrac{g_mR_E}{\alpha}+\dfrac{R_E}{r_o}}\right)} \) \\
    Neglecting Output Resistance \linebreak $\left(r_o \rightarrow \infty\right)$ & \vspace{2pt} \includegraphics[scale=0.4]{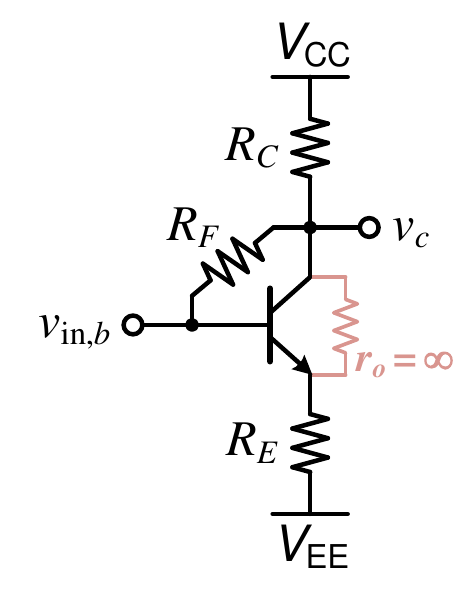} & \(\displaystyle A_v^{\mathrm{CE}} = -\left(\dfrac{g_m}{1+g_m R_E / \alpha}-\dfrac{1}{R_F}\right) \left(R_C \parallel R_F \right) \) \\
    \end{NiceTabular}
\end{table*}

\begin{figure*}[h]
\centering
\subfloat[]{\includegraphics[width=0.5\linewidth]{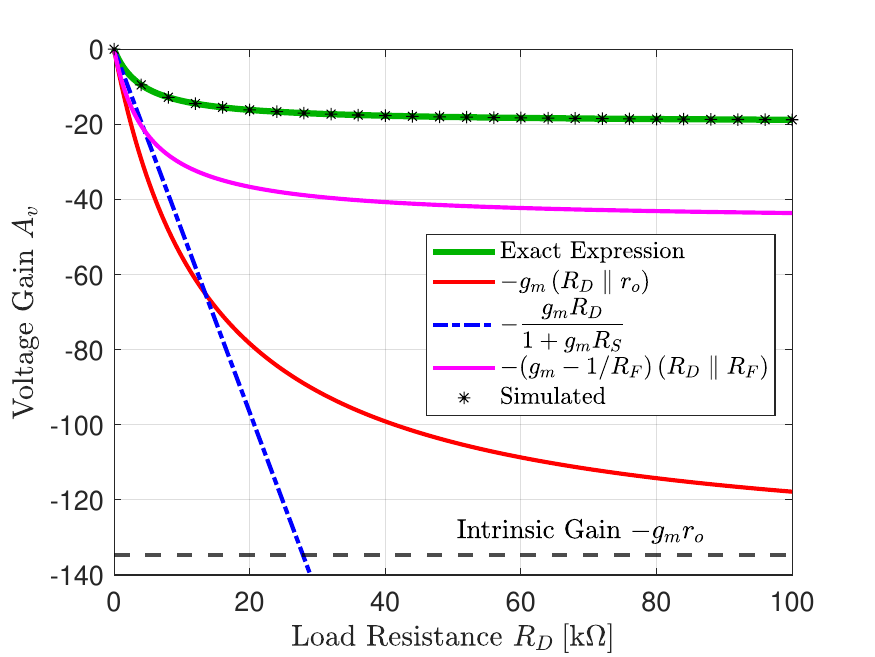}}
\hfill
\subfloat[]{\includegraphics[width=0.5\linewidth]{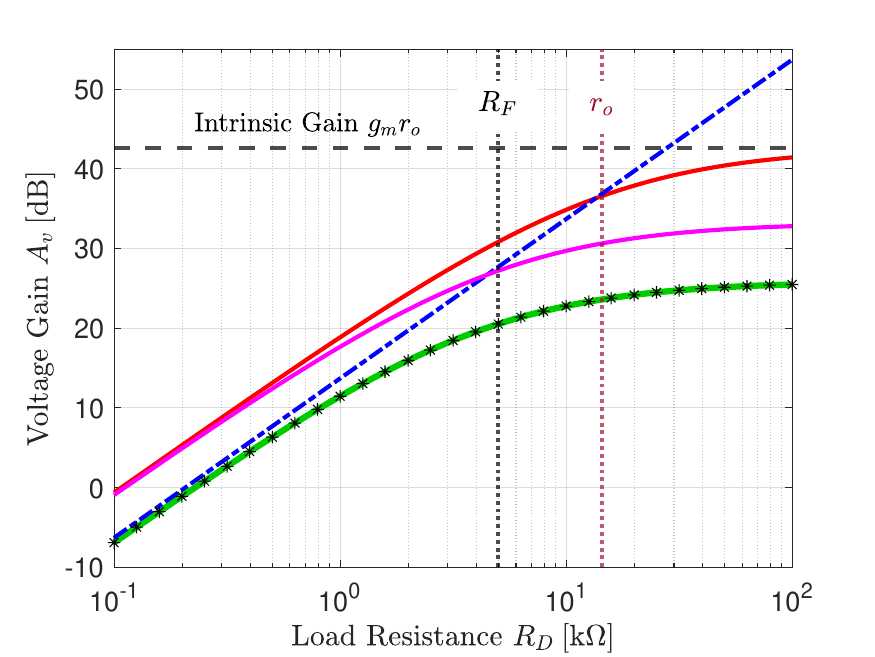}}
\caption{Small-signal voltage gain of a common source amplifier as the load resistance $R_D$ is swept, where $R_E = 100\:\Omega$ and $R_F = 5\:\mathrm{k}\Omega$. The transistor's small-signal parameters were obtained from simulation to be $g_m = 9.36\:\mathrm{mS}$ and $r_o = 14.4\:\mathrm{k}\Omega$. (a) Linear scale. (b) Logarithmic scale, with the gain expressed in decibels.}
\label{fig:CE-Gain-Comp}
\end{figure*}

Fig.~\ref{fig:CE-Gain-Comp} shows the Spectre AC-simulated voltage gain of a common source amplifier constructed from a 36-nm \emph{n}-channel FinFET. Plotted alongside are the expression from the first row of Table~\ref{tab:MOS} as well as several familiar approximations.\footnote{The body effect is neglected here for simplicity.} Note that the simulation is somewhat idealized, and the theory-simulation comparison is thus a bit contrived, for two reasons: AC simulations in Spectre only capture linearized small-signal behavior, and the supply rails of the testbench were constructed so as to keep the transistor's biasing fixed as the load resistance $R_D$ is varied. That said, the empirical validity of the general gain expression, both in an absolute and a comparative sense, is clear. Observe that, for this particular assortment of resistors, (1) the source-degenerated expression is the most accurate for small $R_D$, (2) $R_D = R_F$ marks the point where the feedback approximation becomes more accurate, and (3) $R_D = r_o$ marks the point where $-g_m\left(R_D\parallel r_o\right)$ becomes more accurate.

Lastly, let us consider the feedback-less case of this amplifier (i.e., $R_F\rightarrow\infty$) . Further assume that \hbox{$R_E \ll \beta r_o$} and \hbox{$g_m r_o \gg 1$}---both reasonable approximations. Interestingly, in this scenario, the impact of emitter degeneration depends on the relative sizes of $r_o$ and $R_C$:\footnote{The second approximation relies on the fact that, for all positive $g_mR_E$,
\begin{equation*}
    1 < \left(\cfrac{1+\cfrac{g_mR_E}{\alpha}}{1+\cfrac{g_mR_E}{\beta}}\right) < \beta+1.
\end{equation*}}
\begin{equation*}
\begin{aligned}
\left.A_v^{\mathrm{CE}}\right|_{R_F\rightarrow\infty} &\approx -g_m \left(\dfrac{R_C}{1+g_mR_E/\alpha} \Biggm\Vert \dfrac{r_o}{1+g_mR_E/\beta}\right) \\ \\
&\approx \begin{cases}
-\cfrac{g_mR_C}{1+\cfrac{g_mR_E}{\alpha}}, &r_o \gg R_C \\ \\
-\cfrac{g_mr_o}{1+\cfrac{g_mR_E}{\beta}}, &R_C \gg \left(\beta+1\right) r_o.
\end{cases}
\end{aligned}
\end{equation*}

\section{Common Base Amplifier}
Consider the common base amplifier whose schematic is shown in the first row of Table~\ref{tab:CB}. The input $v_{\mathrm{in},e}$ is applied at the emitter, and the output $v_c$ is taken from the collector. The nodal equations are
\begin{equation}
\begin{aligned}
\begin{pmatrix}
\dfrac{1}{R_C} + \dfrac{1}{R_F} + \dfrac{1}{r_o} & g_m - \dfrac{1}{R_F} \\[1em]
-\dfrac{1}{R_F} & \dfrac{1}{r_{\pi}} + \dfrac{1}{R_B} + \dfrac{1}{R_F}
\end{pmatrix}
\begin{pmatrix}
v_c \\[0.5em] v_b
\end{pmatrix}
\\[1em] =
\begin{pmatrix}
g_m + \dfrac{1}{r_o} \\[1em]
\dfrac{1}{r_{\pi}}
\end{pmatrix}
v_{\mathrm{in},e}.
\end{aligned}
\label{eq:CB-nodal}
\end{equation}
Solving for $A_v^{\mathrm{CB}} \coloneqq v_c/v_{\mathrm{in},e}$ yields the general result in the first row of Table~\ref{tab:CB}. Various approximations of interest are provided in the subsequent rows; the gain of the common gate amplifier is given in the second row of Table~\ref{tab:MOS}.

\begin{table*}[ht]
\centering
\caption{Common Base Amplifier -- Voltage Gain}
\label{tab:CB}
    \begin{NiceTabular}[width=\linewidth]{X[13,c,m] X[15,c,m] X[33,c]}[hvlines]
    \RowStyle[cell-space-limits=6pt]{} General Expression & \includegraphics[scale=0.4]{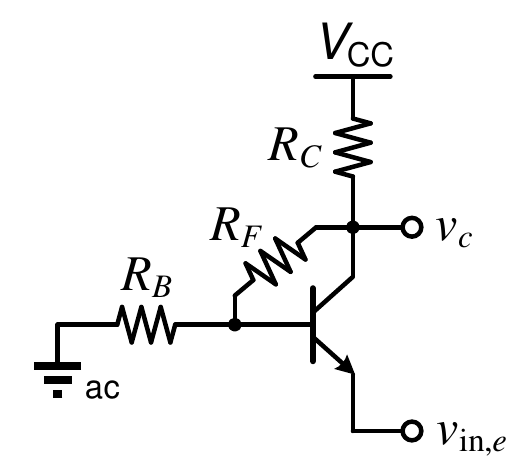} & \(\displaystyle A_v^{\mathrm{CB}} = \dfrac{g_m\left(\dfrac{1+\dfrac{R_B}{\alpha R_F}}{1+\dfrac{g_mR_B}{\beta}+\dfrac{R_B}{R_F}}\right)+\dfrac{1}{r_o}}{\dfrac{1}{R_C \parallel r_o} + \dfrac{1}{R_F}\left(\dfrac{1+\dfrac{g_m R_B}{\alpha}}{1+\dfrac{g_mR_B}{\beta}+\dfrac{R_B}{R_F}}\right)} \) \\
    No Base Degeneration \linebreak $\left(R_B=0\right)$ & \includegraphics[scale=0.4]{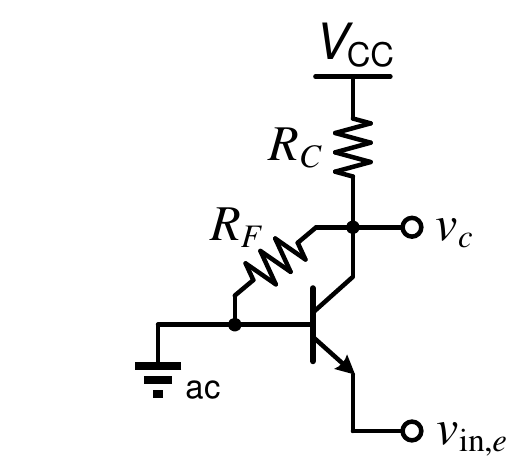} & \(\displaystyle A_v^{\mathrm{CB}} = \left(g_m+\dfrac{1}{r_o}\right) \left(R_C \parallel R_F \parallel r_o\right) \) \\
    No ``Feedback'' \linebreak $\left(R_F \rightarrow \infty\right)$ & \includegraphics[scale=0.4]{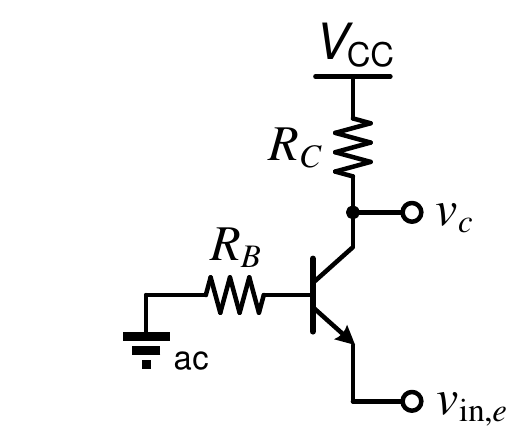} & \(\displaystyle A_v^{\mathrm{CB}} = \left(\dfrac{g_m}{1+g_m R_B / \beta}+\dfrac{1}{r_o}\right) \left(R_C \parallel r_o \right) \) \\
    \RowStyle[cell-space-limits=6pt]{} Neglecting Output Resistance \linebreak $\left(r_o \rightarrow \infty\right)$ & \includegraphics[scale=0.4]{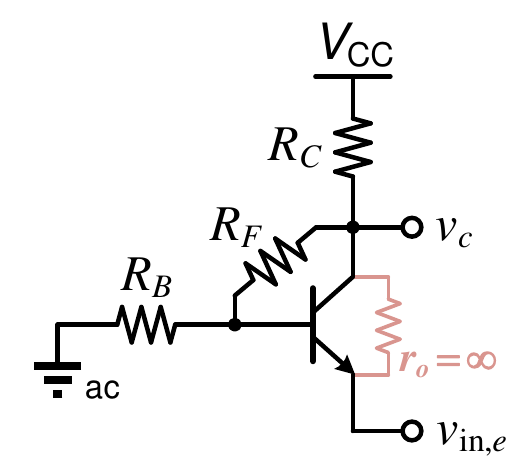} & \(\displaystyle A_v^{\mathrm{CB}} = \dfrac{g_m\left(\dfrac{1+\dfrac{R_B}{\alpha R_F}}{1+\dfrac{g_mR_B}{\beta}+\dfrac{R_B}{R_F}}\right)}{\dfrac{1}{R_C} + \dfrac{1}{R_F}\left(\dfrac{1+\dfrac{g_m R_B}{\alpha}}{1+\dfrac{g_mR_B}{\beta}+\dfrac{R_B}{R_F}}\right)} \) \\
    \end{NiceTabular}
\end{table*}

\begin{figure}[h]
\centering
\includegraphics[width=\columnwidth]{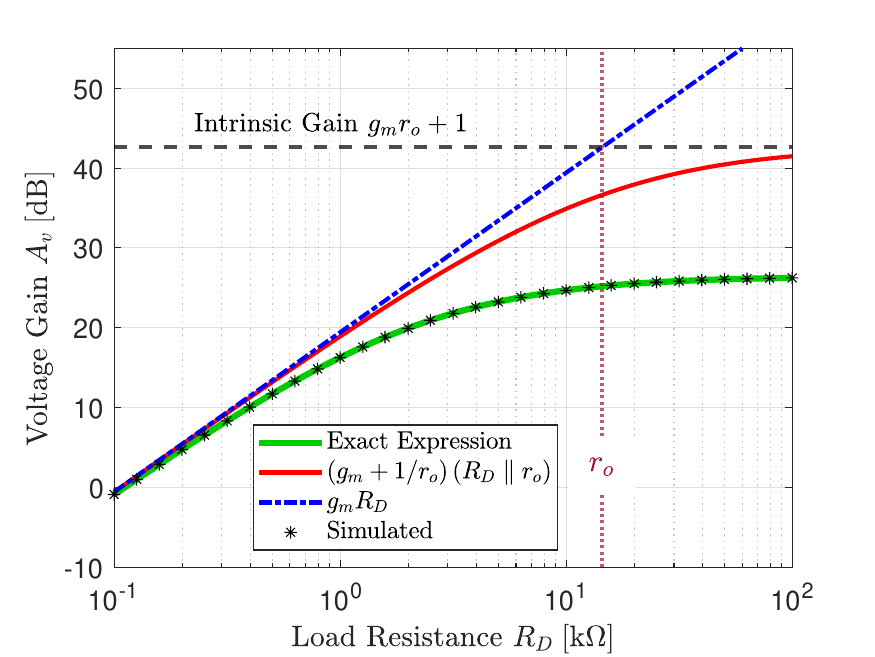}
\caption{Small-signal voltage gain of a common gate amplifier as the drain resistance $R_D$ is varied, where $R_G = 100\:\Omega$, $R_F = 5\:\mathrm{k}\Omega$, and the transistor's small-signal parameters are $g_m = 9.36\:\mathrm{mS}$ and $r_o = 14.4\:\mathrm{k}\Omega$.}
\label{fig:CB-Gain-Comp}
\end{figure}

Fig.~\ref{fig:CB-Gain-Comp} shows the Spectre AC-simulated voltage gain of a common gate amplifier constructed from a 36-nm \emph{n}-channel FinFET. Plotted alongside are the expression from the second row of Table~\ref{tab:MOS} as well as several familiar approximations.\footnote{The body effect is also neglected here.}

\subsection[Base-Emitter Reciprocity]{\texorpdfstring{Base-Emitter Reciprocity Between $A_v^{\mathrm{CE}}$ and $A_v^{\mathrm{CB}}$}{Base-Emitter Reciprocity}}
\label{subsec:cb-recip}
Looking at the first rows of Tables~\ref{tab:CE} and \ref{tab:CB}, there is a striking structural similarity between the common-emitter and common-base gains. This is not a coincidence. For example, the term \hbox{$1+g_mR_B/\beta+R_B/R_F$} serves as a ``base-degeneration'' factor that degrades the transconductance $g_m$ within the expression for $A_v^{\mathrm{CB}}$, in much the same way that the emitter-degeneration term \hbox{$1+g_mR_E/\alpha+R_E/r_o$} degrades $g_m$ within the expression for $A_v^{\mathrm{CE}}$. More specifically, observe how $r_o$ and $R_F$ appear to have ``switched'' places between $A_v^{\mathrm{CE}}$ and $A_v^{\mathrm{CB}}$. This is unsurprising: $r_o$ now forms a feedback path between the common base amplifier's input and output (just like $R_F$ does for the common emitter amplifier), whereas $R_F$ connects the output to the common (base) terminal (the very role played by $r_o$ in the common-emitter scenario).\footnote{Calling $R_F$ a ``feedback'' resistor here is a bit of a misnomer, but, for the sake of symbolic consistency, we will not re-label that resistance.} In other words, these two resistances have indeed ``swapped'' positions with each other in terms of how they affect the circuit's transfer characteristic. In addition, notice how $\alpha$ and $\beta$ also ``switched'' places, a reflection of how the two amplifiers' input impedances \emph{roughly} relate to one another: $r_{\pi} = \beta/g_m$ looking into the base versus $r_e = \alpha/g_m$ looking into the emitter.\footnote{A rigorous treatment of the transistor's terminal impedances follows in Sections~\ref{sec:rb}-\ref{sec:rc}.}

Consequently,  the common base gain also features an ``intuitive'' interpretation analogous to \eqref{eq:CE_intuitive}:
\begin{equation}
A_v^{\mathrm{CB}} = \left(g_{m,\mathrm{eff}}^{\mathrm{CB}} + 1/r_o\right) \left(R_C \parallel R_{F,\mathrm{eff}}^{\mathrm{CB}} \parallel r_o\right),
\label{eq:CB_intuitive}
\end{equation}
where
\begin{equation*}
g_{m,\mathrm{eff}}^{\mathrm{CB}} = g_m \left(\dfrac{1+\dfrac{R_B}{\alpha R_F}}{1+\dfrac{g_mR_B}{\beta}+\dfrac{R_B}{R_F}}\right)
\end{equation*}
and
\begin{equation*}
R_{F,\mathrm{eff}}^{\mathrm{CB}} = R_F \left(\dfrac{1+\dfrac{g_mR_B}{\beta}+\dfrac{R_B}{R_F}}{1+\dfrac{g_m R_B}{\alpha}}\right).
\end{equation*}

\begin{figure*}[h]
\subfloat[]{\includegraphics[scale=0.34]{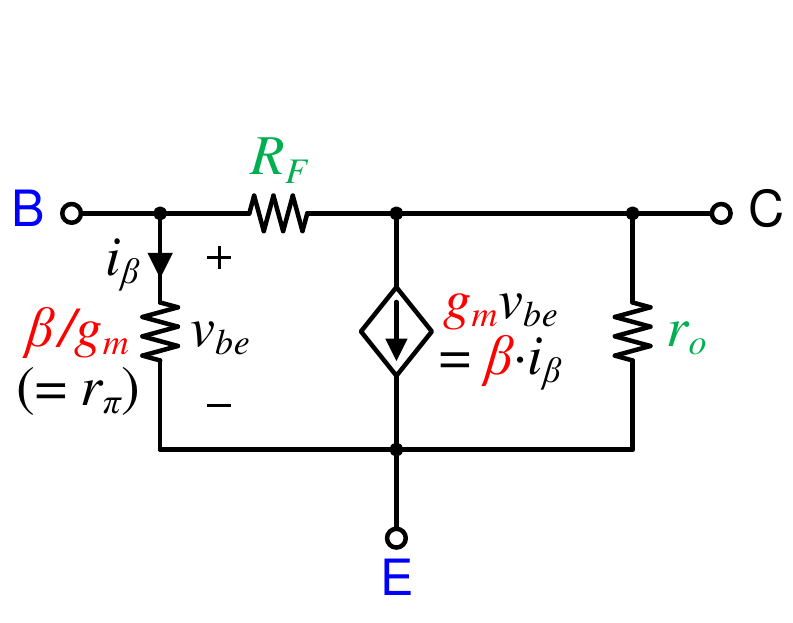}}
\hfill
\subfloat[]{\includegraphics[scale=0.34]{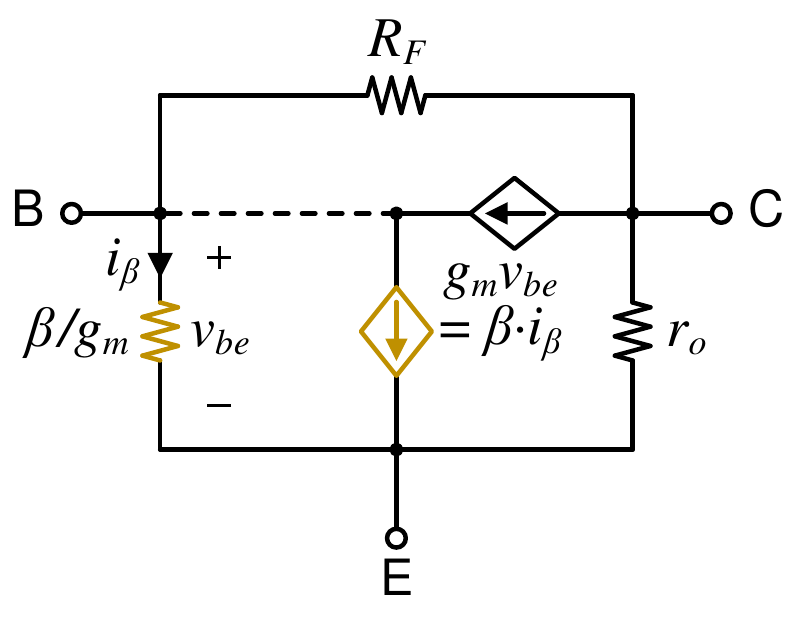}}
\hfill
\subfloat[]{\includegraphics[scale=0.34]{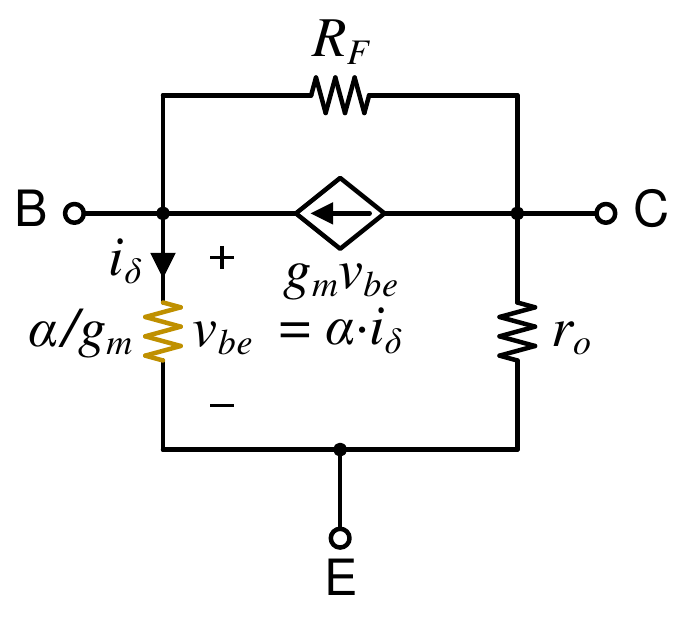}}
\hfill
\subfloat[]{\includegraphics[scale=0.34]{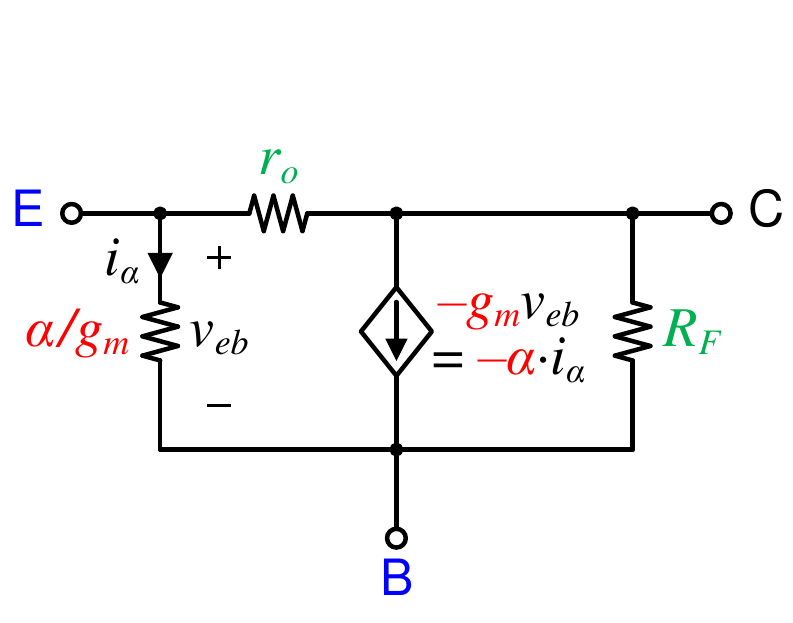}}
\caption{Illustrating base-emitter reciprocity through a sequence of equivalent small-signal models. (a) Starting with the $\pi$-model, but including a feedback resistor $R_F$. (b) Duplicating the dependent current source and noting that the dashed connection does not alter the circuit's behavior because no current would flow across it. (c) The dashed connection turns the gold-colored dependent current source into a resistance of $1/g_m$, which can be combined in parallel with the gold-colored $\beta/g_m$ resistor. (d) Rearrangement in the ``form'' of the $\pi$-model of (a), thereby revealing the desired base-emitter reciprocity relation.}
\label{fig:BE-recip}
\end{figure*}

We can make the foregoing observations rigorous by transforming the small-signal $\pi$-model of the transistor into an equivalent form where the base and emitter terminals have traded positions, as shown in Fig.~\ref{fig:BE-recip}. While the technical details of the transformation are supplied in the figure's caption\footnote{Note that Figs.~\ref{fig:BE-recip}(b) and (c) essentially depict the well-known steps for converting the $\pi$-model to the $T$-model \cite{Ali}.}, we can compare Figs.~\ref{fig:BE-recip}(a) and (d) to surmise that ``swapping'' $g_m$ with $-g_m$, $\alpha$ with $-\beta$, and $r_o$ with $R_F$ is equivalent to exchanging the base and emitter terminals. Therefore, upon further exchanging the loads at these terminals, $R_E$ and $R_B$, these four swaps,
\begin{equation}
\begin{alignedat}{2}
g_m &\Longleftrightarrow -g_m \qquad \alpha &&\Longleftrightarrow -\beta \\ r_o &\Longleftrightarrow R_F \qquad R_E &&\Longleftrightarrow R_B,
\end{alignedat}
\label{eq:BE-recip}
\end{equation}
allow us to mathematically convert between the common emitter and common base topologies---including their underlying nodal equations [i.e., \eqref{eq:CE-nodal} and \eqref{eq:CB-nodal}] as well as their gain and impedance expressions. We call this principle ``base-emitter reciprocity.'' We will say that two quantities, $x$ and $y$, are ``base-emitter reciprocals'' of one another if applying the set of transformations of \eqref{eq:BE-recip} converts $x$ into $y$ (and vice versa), and we shall denote this as $x \rightleftharpoons y$. 

With this reciprocity concept formalized, we can apply it to the common-emitter and common-base gain expressions (Tables~\ref{tab:CE} and \ref{tab:CB}) to write down the following:
\begin{equation}
\begin{alignedat}{2}
&A_v^{\mathrm{CE}} \quad &&\rightleftharpoons \quad A_v^{\mathrm{CB}} \\
&A_v^{\mathrm{CE}}\left(R_E = 0\right) \quad &&\rightleftharpoons \quad A_v^{\mathrm{CB}}\left(R_B = 0\right) \\
&A_v^{\mathrm{CE}}\left(R_F\rightarrow\infty\right) \quad &&\rightleftharpoons \quad A_v^{\mathrm{CB}}\left(r_o\rightarrow\infty\right) \\
&A_v^{\mathrm{CE}}\left(r_o\rightarrow\infty\right) \quad &&\rightleftharpoons \quad A_v^{\mathrm{CB}}\left(R_F\rightarrow\infty\right)
\end{alignedat}
\end{equation}

\section[Common Collector Amplifier (Emitter Follower)]{Common Collector Amplifier \\ (Emitter Follower)}
Consider the common collector amplifier whose schematic is shown in the first row of Table~\ref{tab:CC}. The nodal equations are the same as those for the common emitter amplifier \eqref{eq:CE-nodal}, but the output $v_e$ is taken from the emitter instead. Solving for $A_v^{\mathrm{CC}} \coloneqq v_e/v_{\mathrm{in},b}$ yields the general result in the first row of Table~\ref{tab:CC}. Various approximations of interest are provided in the subsequent rows; the gain of the common drain amplifier is given in the third row of Table~\ref{tab:MOS}. The common collector amplifier is also known as an ``emitter follower''---so named because the emitter voltage roughly \emph{follows} the input voltage. One can deduce that the voltage gain is typically close to unity so long as $g_mR_E$ is sizable.

\begin{table*}[ht]
\centering
\caption{Common Collector Amplifier (Emitter Follower) -- Voltage Gain}
\label{tab:CC}
    \begin{NiceTabular}[width=\linewidth]{X[13,c,m] X[15,c,m] X[33,c]}[hvlines]
    General Expression & \vspace{2pt} \includegraphics[scale=0.4]{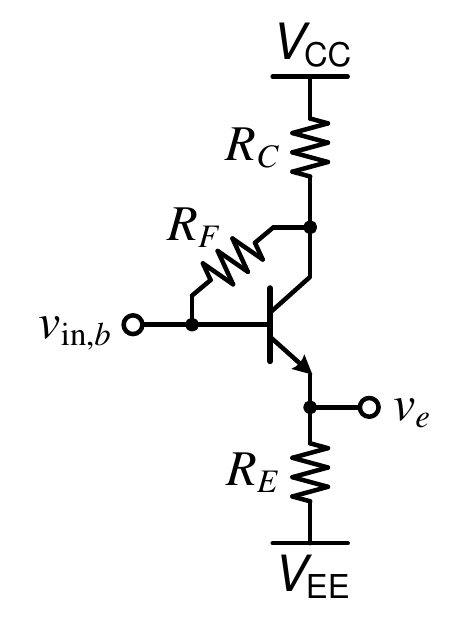} & \(\displaystyle A_v^{\mathrm{CC}} = \dfrac{\dfrac{g_m}{\alpha}r_o + \dfrac{g_m}{\beta}\left(R_C \parallel R_F\right) + \dfrac{R_C}{R_C+R_F}}{\dfrac{g_m}{\alpha}r_o + \dfrac{g_m}{\beta}\left(R_C \parallel R_F\right) + \dfrac{r_o + \left(R_C \parallel R_F\right)}{R_E} + 1} \) \\
    Current Source Load \linebreak $\left(R_E\rightarrow\infty\right)$ & \vspace{2pt} \includegraphics[scale=0.4]{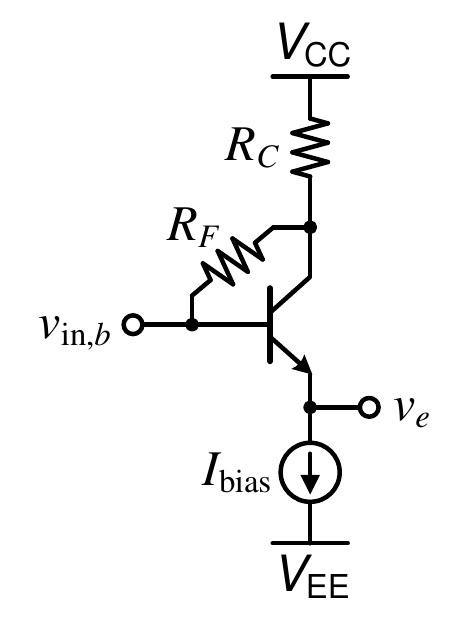} & \(\displaystyle A_v^{\mathrm{CC}} = \dfrac{\dfrac{g_m}{\alpha}r_o + \dfrac{g_m}{\beta}\left(R_C \parallel R_F\right) + \dfrac{R_C}{R_C+R_F}}{\dfrac{g_m}{\alpha}r_o + \dfrac{g_m}{\beta}\left(R_C \parallel R_F\right) + 1} \) \\
    Diode-Connected \linebreak $\left(R_F = 0\right)$ & \vspace{2pt} \includegraphics[scale=0.4]{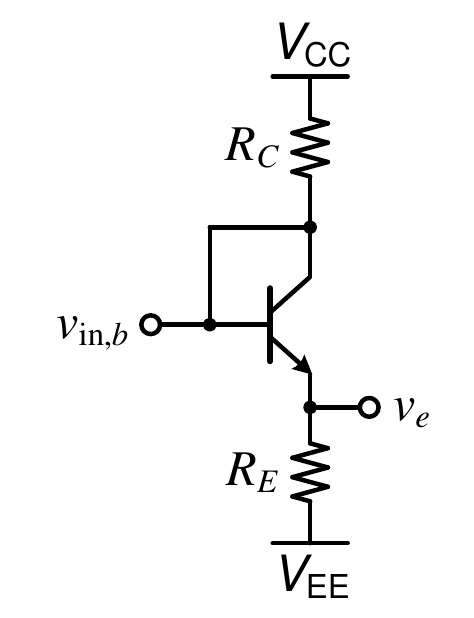} & \(\displaystyle A_v^{\mathrm{CC}} = \dfrac{\dfrac{g_m}{\alpha} \left(R_E \parallel r_o\right) + \dfrac{R_E}{R_E + r_o}}{\dfrac{g_m}{\alpha} \left(R_E \parallel r_o\right) + 1} \) \\
    Collector Shorted \linebreak to Supply \linebreak $\left(R_C = 0\right)$ & \vspace{2pt} \includegraphics[scale=0.4]{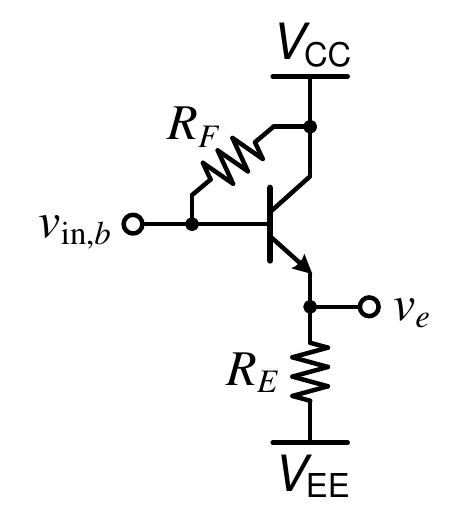} & \(\displaystyle A_v^{\mathrm{CC}} = \dfrac{\dfrac{g_m}{\alpha} \left(R_E \parallel r_o\right)}{\dfrac{g_m}{\alpha} \left(R_E \parallel r_o\right) + 1} \) \\
    Neglecting Output Resistance \linebreak $\left(r_o \rightarrow \infty\right)$ & \vspace{2pt} \includegraphics[scale=0.4]{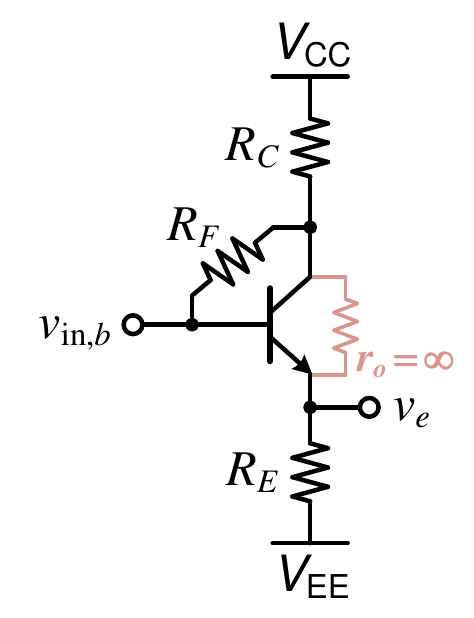} & \(\displaystyle A_v^{\mathrm{CC}} = \dfrac{g_mR_E/\alpha}{g_mR_E/\alpha + 1} \) \\
    \end{NiceTabular}
\end{table*}

The diode-connected case ($R_F = 0$) is included mostly to illustrate that its gain can be simplified as
\begin{equation*}
A_v^{\mathrm{CC}}\!\left(R_F=0\right) = \frac{R_E}{R_E + \left(\alpha r_m \parallel r_o\right)},
\end{equation*}
which is a voltage divider between $R_E$ and the small-signal resistance across the diode-connected transistor, $\left(\alpha r_m \parallel r_o\right)$. Practically speaking, its utility as a voltage buffer is severely limited by its very low input impedance.\footnote{Were this not true, why not just replace the diode-connected transistor with an actual diode, leading to a gain of $R_E/\left(R_E + r_d\right)$ where $r_d$ is the diode's small-signal resistance?} We can see this by contrasting the emitter follower's input impedance $r_{\mathrm{in}}^{\mathrm{CC}}$ between the $R_F = 0$ and $R_F = \infty$ cases. To keep things simple, assume that $r_o \gg R_C \gg R_E$. If $R_F\rightarrow\infty$, then\footnote{See Section~\ref{sec:rb} and Table~\ref{tab:RB}.}
\begin{equation*}
\begin{aligned}
r_{\mathrm{in}}^{\mathrm{CC}}\!\left(R_F\rightarrow\infty\right) &= r_{\pi} + R_E\left[\dfrac{\left(\beta+1\right)r_o + R_C}{r_o+R_C+R_E}\right] \\ &\approx r_{\pi} + \left(\beta+1\right)R_E \\ &= \left(\beta+1\right) \left(\alpha r_m + R_E\right),
\end{aligned}
\end{equation*}
On the other hand, if the transistor is diode-connected, it is easy to see that
\begin{equation*}
\begin{aligned}
r_{\mathrm{in}}^{\mathrm{CC}}\!\left(R_F=0\right) &= R_C \parallel \left[R_E + \left(\alpha r_m \parallel r_o\right)\right] \\ &\approx \alpha r_m + R_E.
\end{aligned}
\end{equation*}
Diode-connecting the transistor therefore reduced the input impedance by a factor of $\left(\beta+1\right)$.\footnote{A numerical example: suppose $g_m = 10\:\mathrm{mS}$, $\beta = 100$, $r_o = 10\:\mathrm{k}\Omega$, $R_C = 1\:\mathrm{k}\Omega$, and $R_E = 100\:\Omega$. While the gain barely changes---$A_v^{\mathrm{CC}}\left(R_F\rightarrow\infty\right) = 0.48$ vs.\ $A_v^{\mathrm{CC}}\left(R_F=0\right) = 0.5$---the input impedance drops by more than a hundredfold, from $r_{\mathrm{in}}^{\mathrm{CC}}\left(R_F\rightarrow\infty\right) = 19.1\:\mathrm{k}\Omega$ to $r_{\mathrm{in}}^{\mathrm{CC}}\left(R_F=0\right) = 165\:\Omega$.}

\begin{figure*}[h]
\centering
\subfloat[]{\includegraphics[width=0.5\linewidth]{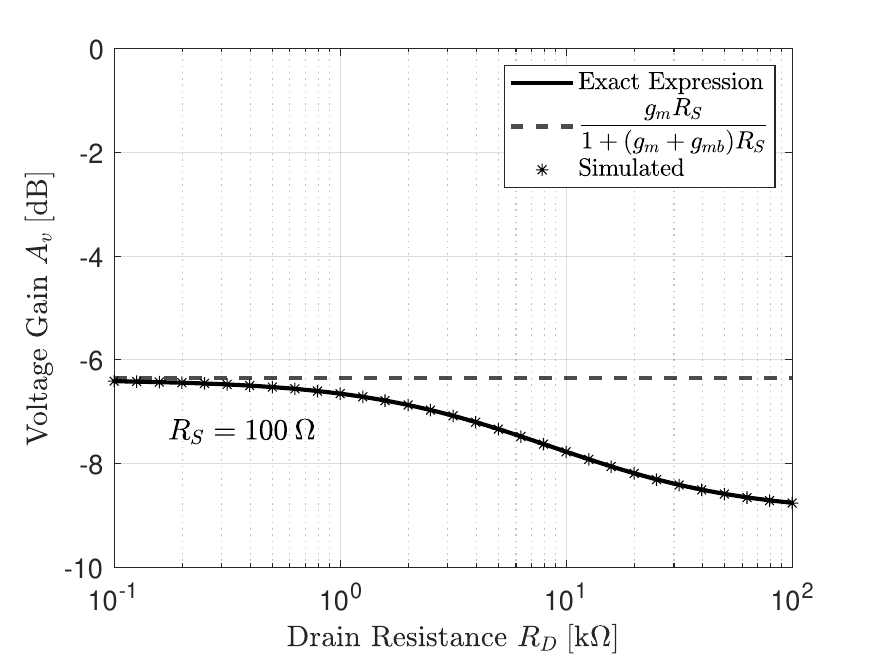}}
\hfill
\subfloat[]{\includegraphics[width=0.5\linewidth]{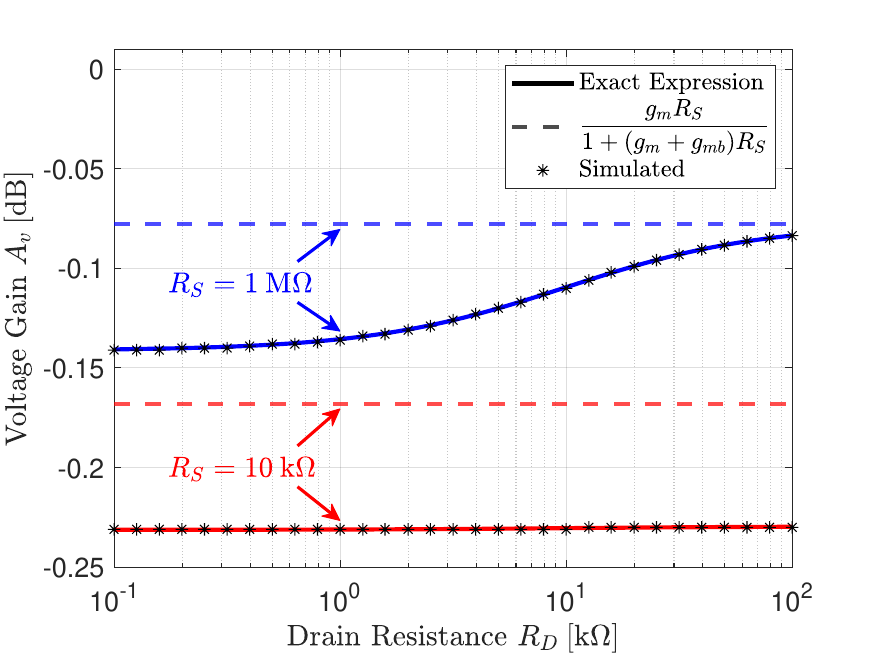}}
\caption{Small-signal voltage gain of a source follower as the drain resistance $R_D$ is varied, where $R_F = 5\:\mathrm{k}\Omega$ and the transistor's small-signal parameters are $g_m = 9.36\:\mathrm{mS}$, $g_{mb} = 83.1\:\mu\mathrm{S}$, and $r_o = 14.4\:\mathrm{k}\Omega$. Results for three different loads are shown: (a) $R_S = 100\:\Omega$. (b) $R_S = 10\:\mathrm{k}\Omega$ and $R_S = 1\:\mathrm{M}\Omega$.}
\label{fig:CC-Gain-Comp}
\end{figure*}

Simulation results for a common drain amplifier constructed from a 36-nm \emph{n}-channel FinFET are shown in Fig.~\ref{fig:CC-Gain-Comp}, plotted alongside the theoretical expression from the third row of Table~\ref{tab:MOS}. The familiar approximation that neglects $r_o$ (and therefore does not vary with $R_D$) is also depicted. The body effect is included in this particular set of simulations only to show how even the minute degree to which it affects the gain is accurately captured by our exact expression. As expected, the body effect fundamentally limits how close to unity (or $0\:\mathrm{dB}$) the gain can get through increasing $R_S$.

\section{Impedance Looking into Base}
\label{sec:rb}
Consider the circuit shown in the first row of Table~\ref{tab:RB}, wherein we are interested in obtaining the resistance $r_b$ looking into the base; to do so, we apply a test current $i_T$ at the base and probe the voltage across it, leading to the following nodal equations:
\begin{equation}
\begin{aligned}
\begin{pmatrix}
\dfrac{1}{r_{\pi}} + \dfrac{1}{R_F} & -\dfrac{1}{r_{\pi}} & -\dfrac{1}{R_F} \\[1em]
-g_m - \dfrac{1}{r_{\pi}} & g_m + \dfrac{1}{r_{\pi}} + \dfrac{1}{R_E} + \dfrac{1}{r_o} & -\dfrac{1}{r_o} \\[1em]
g_m - \dfrac{1}{R_F} & -g_m - \dfrac{1}{r_o} & \dfrac{1}{R_C} + \dfrac{1}{R_F} + \dfrac{1}{r_o}
\end{pmatrix}
\\[1em]
\begin{pmatrix}
v_b \\[0.5em] v_e \\[0.5em] v_c
\end{pmatrix}
=
\begin{pmatrix}
1 \\[0.5em] 0 \\[0.5em] 0
\end{pmatrix}
i_T.
\end{aligned}
\label{eq:RB-nodal}
\end{equation}
Solving for $r_b \coloneqq v_b/i_T$ yields the expression in the first row of Table~\ref{tab:RB}. Various approximations of interest are provided in the subsequent rows, and the analogous impedance looking into the gate of a MOSFET is given in the fourth row of Table~\ref{tab:MOS}.

\begin{table*}[ht]
\centering
\caption{Resistance Looking into Base}
\label{tab:RB}
    \begin{NiceTabular}[width=\linewidth]{X[9,c,m] X[13,c,m] X[38,c]}[hvlines]
    General Expression & \includegraphics[scale=0.4]{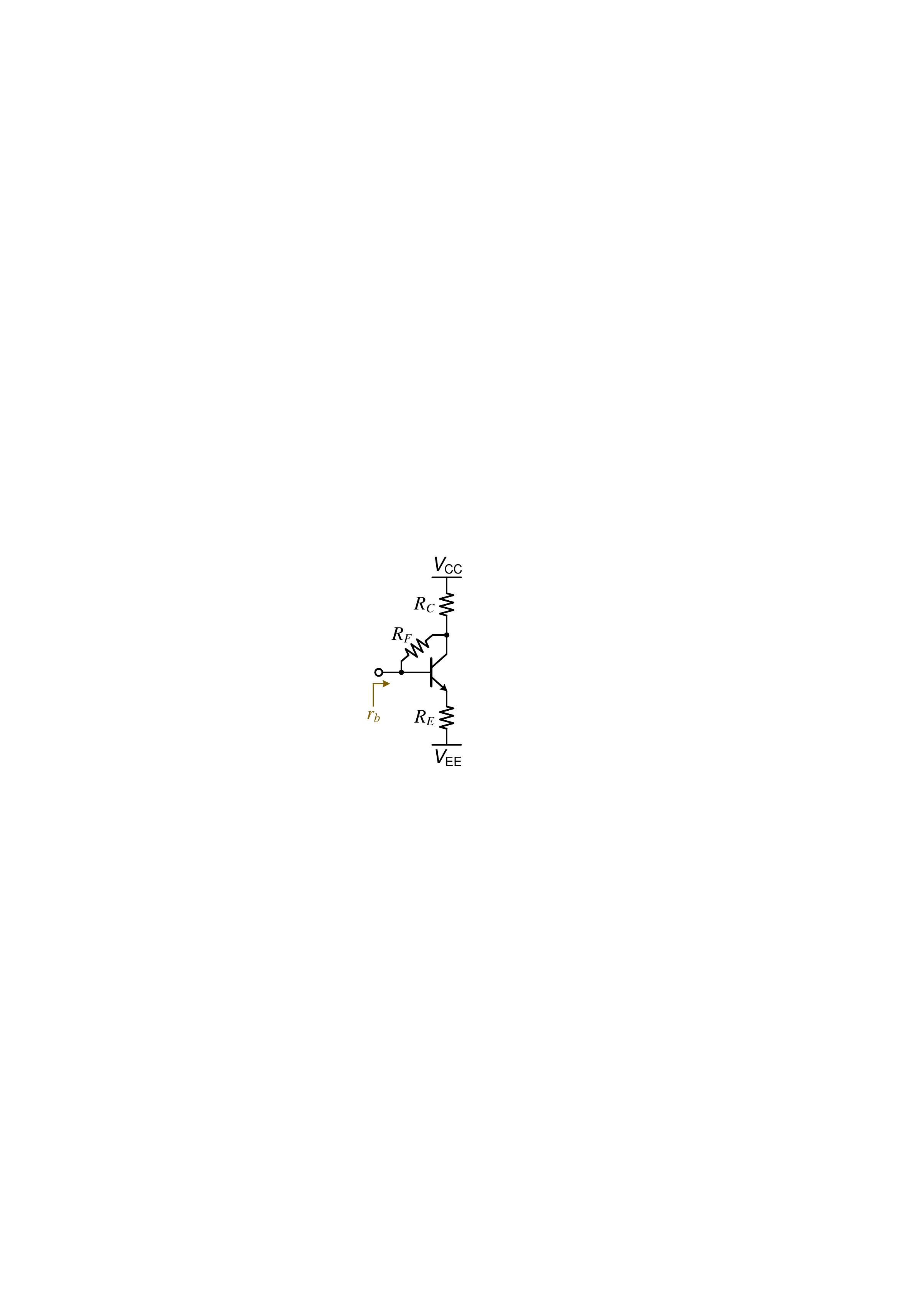} \vspace{-3pt} & \(\displaystyle r_b = \dfrac{R_F + \left[R_C\parallel (R_E+r_o)\right]+ g_m\left(\dfrac{r_o}{\alpha} + \dfrac{R_C \parallel R_F}{\beta}\right) \dfrac{R_E(R_C+R_F)}{R_C+R_E+r_o}}{1 + \dfrac{g_mR_F}{\beta} + \dfrac{g_m\left[(R_C+R_E) \parallel r_o\right]}{\alpha}} \) \\
    No Emitter Degeneration \linebreak $\left(R_E=0\right)$ & \includegraphics[scale=0.4]{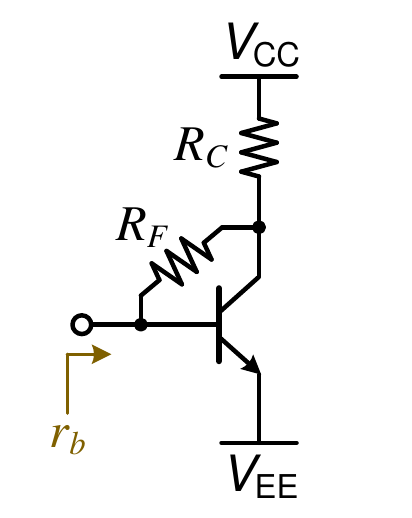} \vspace{-3pt} & \(\displaystyle r_b = r_{\pi} \biggm\Vert \dfrac{R_F + \left(R_C \parallel r_o\right)}{1 + g_m \left(R_C \parallel r_o\right)} = \alpha r_m \biggm\Vert \dfrac{R_F + \left(R_C \parallel r_o\right)}{1 - g_mR_F} \) \\
    No Feedback \linebreak $\left(R_F \rightarrow \infty\right)$ & \includegraphics[scale=0.4]{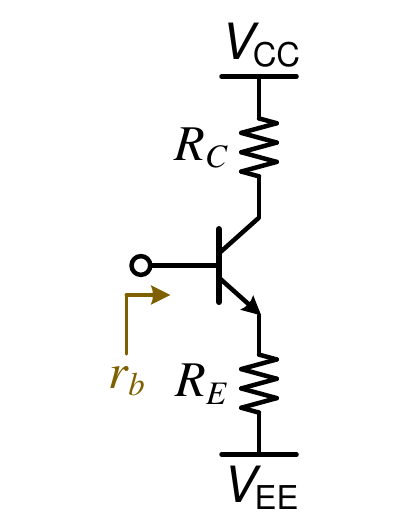} \vspace{-3pt} & \(\displaystyle r_b = r_{\pi} + R_E\left[\dfrac{ \left(\beta+1\right)r_o + R_C }{r_o+R_C+R_E}\right] \) \\
    Neglecting Output Resistance \linebreak $\left(r_o \rightarrow \infty\right)$ & \includegraphics[scale=0.4]{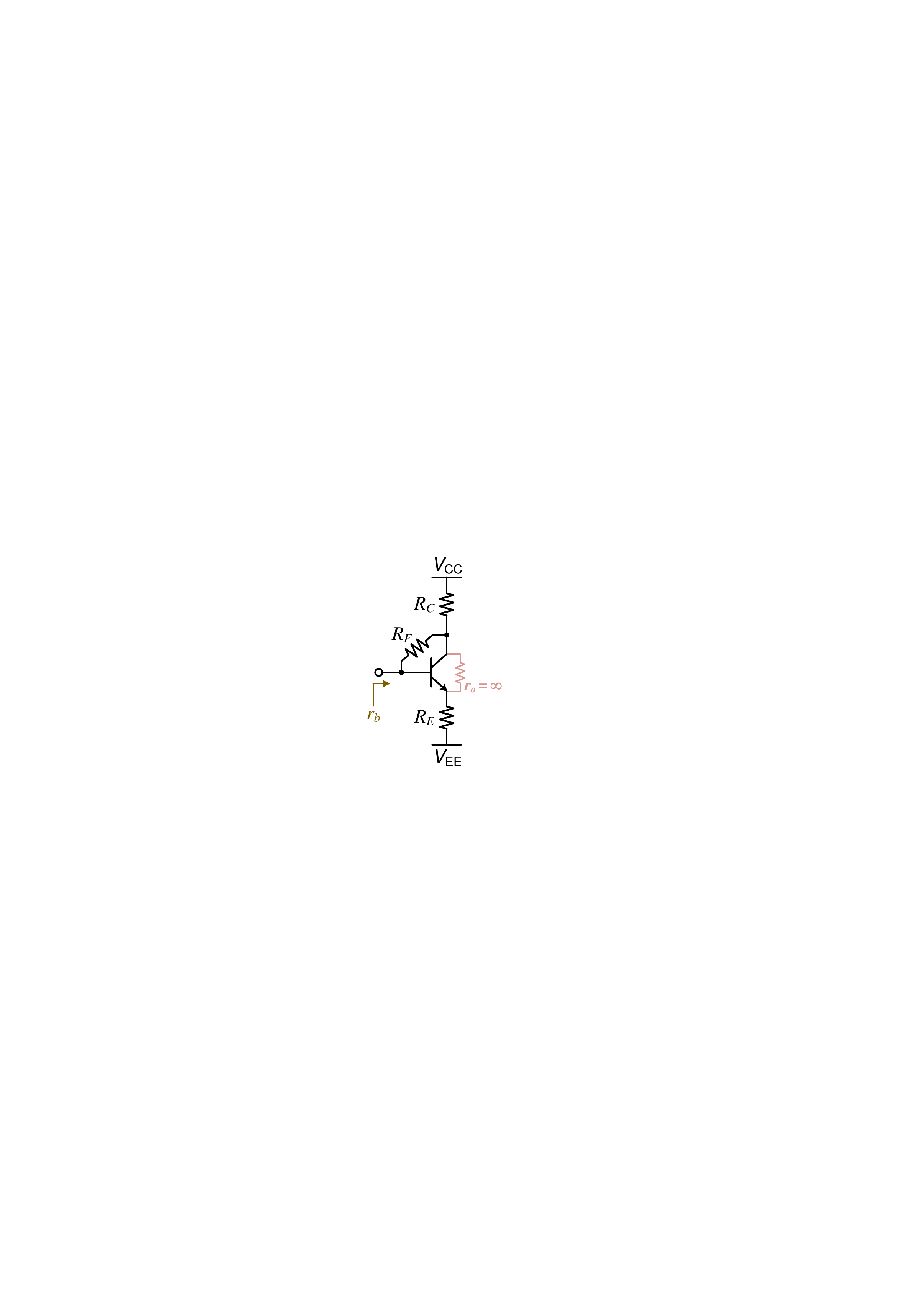} \vspace{-3pt} & \(\displaystyle \begin{aligned} r_b = \left[r_{\pi} + \left(\beta+1\right)R_E\right] &\left\Vert \left( \dfrac{R_C+R_F}{1 + \dfrac{g_mR_C}{1+g_mR_E/\alpha}}\right)\right. \\ = \left(\alpha r_m + R_E\right) &\left\Vert \left( \dfrac{R_C+R_F}{1 - \dfrac{g_mR_F}{1+g_mR_E/\alpha}}\right)\right. \end{aligned} \) \\
    Current-Source Bias at Emitter \linebreak $\left(R_E\rightarrow\infty\right)$ & \includegraphics[scale=0.4]{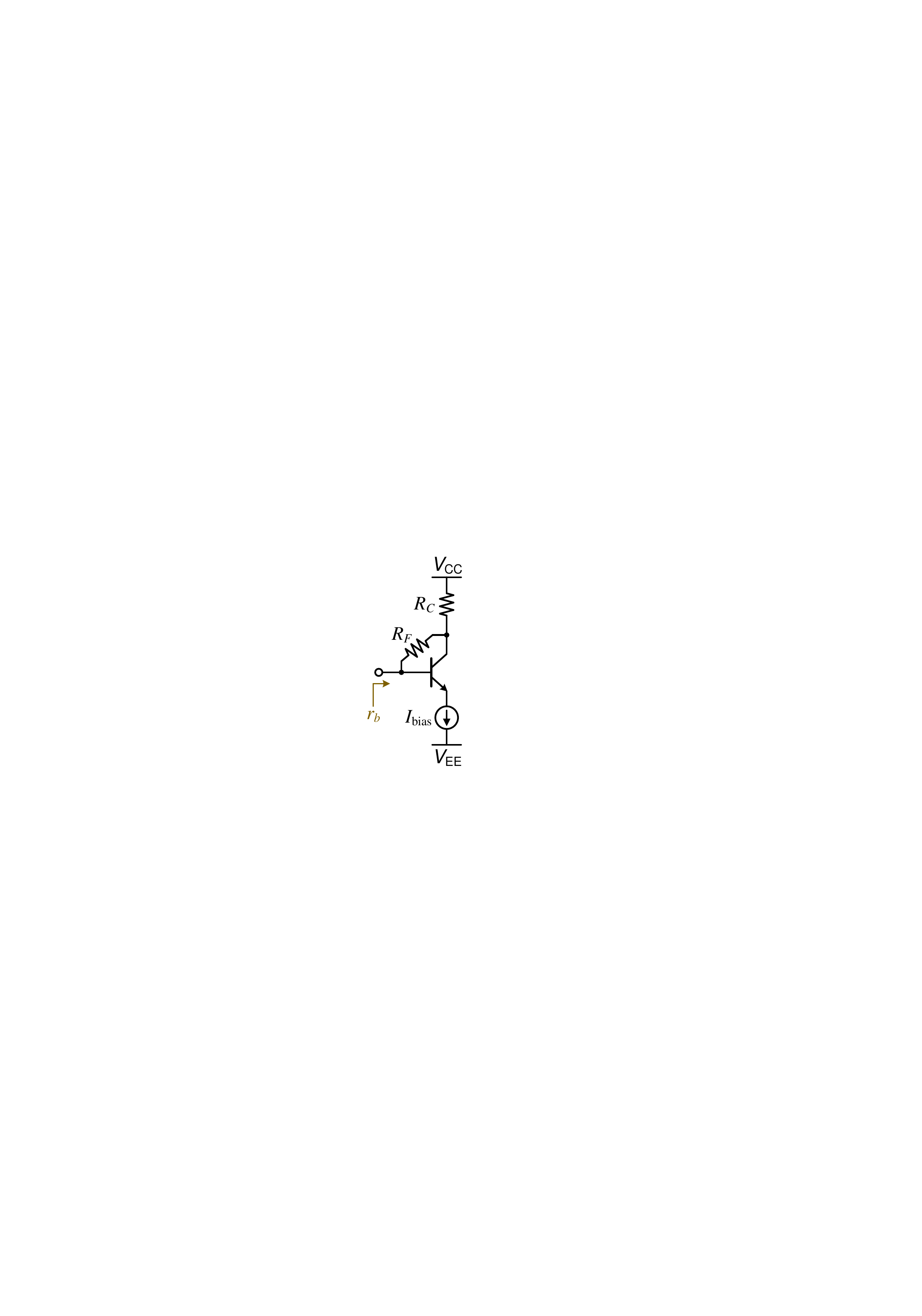} \vspace{-3pt} & \(\displaystyle r_b = \left[\left(r_{\pi} + \left(\beta+1\right)r_o\right) \!\parallel\! R_F\right] + R_C \) \linebreak\linebreak (Compare with $r_c$ assuming $R_E\rightarrow\infty$.) \\
    Current-Source Bias at Collector \linebreak $\left(R_C\rightarrow\infty\right)$ & \includegraphics[scale=0.4]{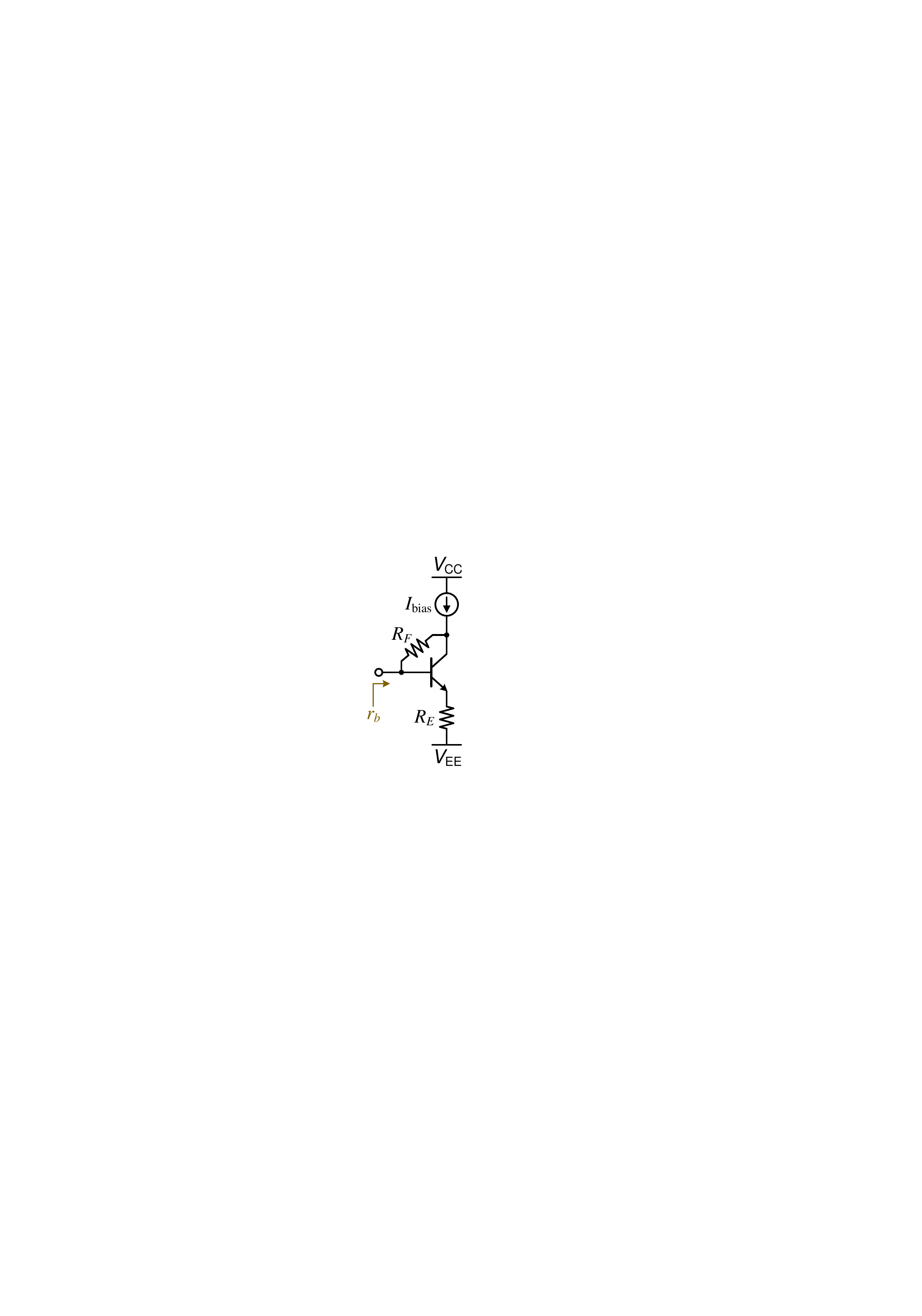} \vspace{-3pt} & \(\displaystyle r_b = \left(r_{\pi} \biggm\Vert \dfrac{R_F+r_o}{1+g_mr_o}\right) + R_E = \left(\alpha r_m \biggm\Vert \dfrac{R_F+r_o}{1-g_mR_F}\right) + R_E \) \linebreak\linebreak\linebreak (Compare with $r_e$ assuming $R_C\rightarrow\infty$.) \\
    \end{NiceTabular}
\end{table*}

Fig.~\ref{fig:RB-RE-Comp}(a) shows the Spectre AC-simulated resistance looking into the gate of a 36-nm \emph{n}-channel FinFET, plotted alongside the expression from Table~\ref{tab:MOS}'s fourth row.\footnote{The body effect is neglected here.} The asymptotic behaviors for $R_D = 0$ and $R_D \rightarrow\infty$ are also indicated.\footnote{The expression for the the $R_D\rightarrow\infty$ asymptote assumes $g_m r_o \gg 1$.}

\section{Impedance Looking into Emitter}
\label{sec:re}
Consider the circuit shown in the first row of Table~\ref{tab:RE}, wherein we are interested in obtaining the resistance $r_e$ looking into the emitter; to do so, we apply a test current $i_T$ at the emitter and probe the voltage across it, leading to the following nodal equations:
\begin{equation}
\begin{aligned}
\begin{pmatrix}
g_m + \dfrac{1}{r_{\pi}} + \dfrac{1}{r_o} & -g_m - \dfrac{1}{r_{\pi}} & -\dfrac{1}{r_o} \\[1em]
-\dfrac{1}{r_{\pi}} & \dfrac{1}{r_{\pi}} + \dfrac{1}{R_B} + \dfrac{1}{R_F} & -\dfrac{1}{R_F} \\[1em]
-g_m - \dfrac{1}{r_o} & g_m - \dfrac{1}{R_F} & \dfrac{1}{R_C} + \dfrac{1}{R_F} + \dfrac{1}{r_o}
\end{pmatrix}
\\[1em]
\begin{pmatrix}
v_e \\[0.5em] v_b \\[0.5em] v_c
\end{pmatrix}
=
\begin{pmatrix}
1 \\[0.5em] 0 \\[0.5em] 0
\end{pmatrix}
i_T.
\end{aligned}
\label{eq:RE-nodal}
\end{equation}
Solving for $r_e \coloneqq v_e/i_T$ yields the expression in the first row of Table~\ref{tab:RE}. Various approximations of interest are provided in the subsequent rows, and the analogous impedance looking into the source of a MOSFET is given in the fifth row of Table~\ref{tab:MOS}. Notice how the impedance looking into the source of an MOS transistor can be written as
\begin{equation*}
    r_s = \dfrac{\left. r_s \right|_{g_{mb}=0}}{1 + g_{mb} \left(r_m \parallel r_o\right)},
\end{equation*}
where
\begin{equation*}
     \left. r_s \right|_{g_{mb}=0} = \left( r_m \parallel r_o\right) + \dfrac{R_D \left(R_G + \dfrac{R_F}{1+g_mr_o}\right)}{R_G + R_D + R_F}
\end{equation*}
is the value of this impedance in the absence of the body effect. This comports with our conventional understanding that the body effect lowers $r_s$.\footnote{If $R_D = 0$ and we neglect $r_o$, we arrive at the familiar result that the body effect lowers $r_s$ from $1/g_m$ to $1/\left(g_m + g_{mb}\right)$.}

\begin{figure*}[h]
\centering
\subfloat[]{\includegraphics[width=0.5\linewidth]{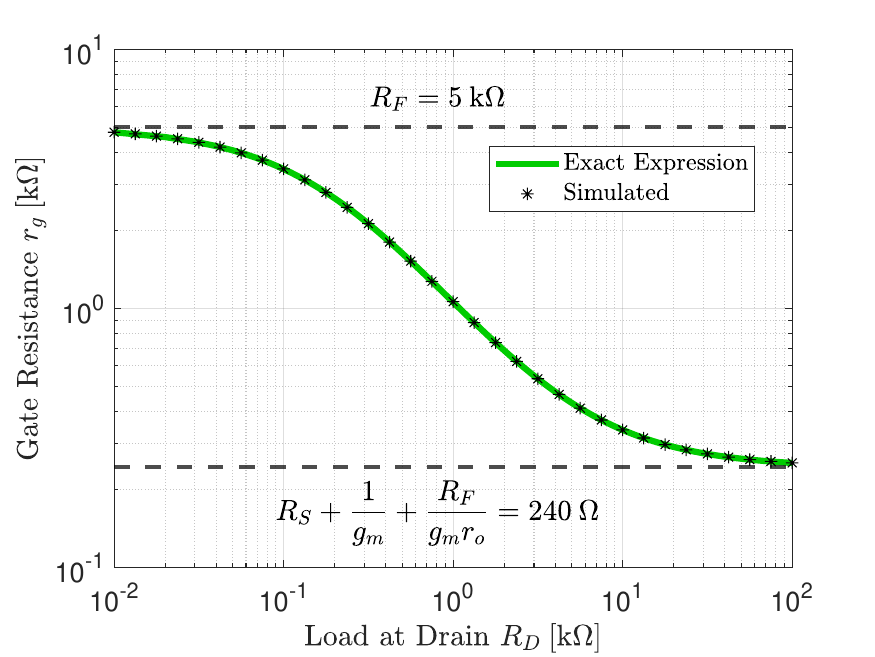}}
\hfill
\subfloat[]{\includegraphics[width=0.5\linewidth]{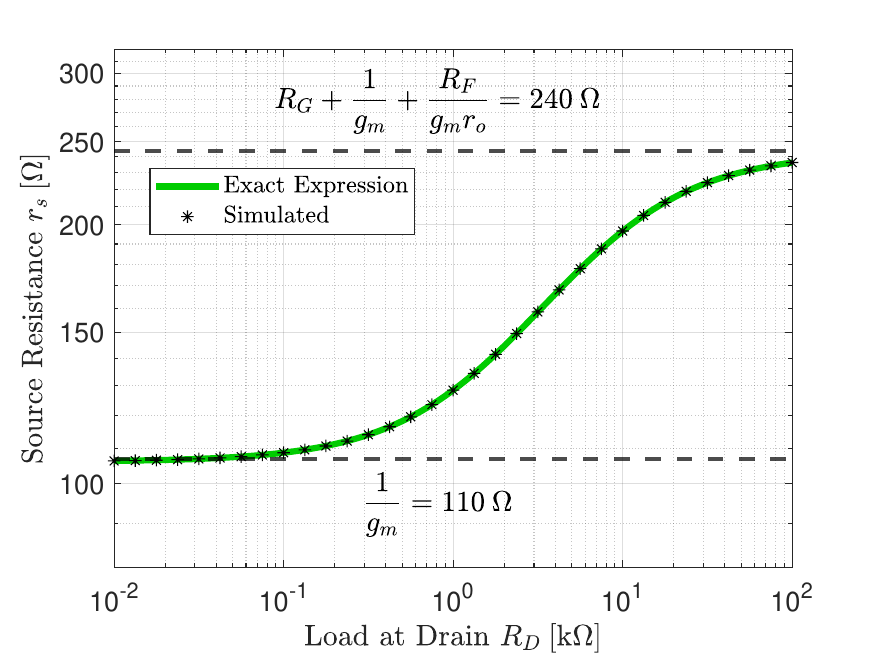}}
\caption{Small-signal resistance looking (a) into the gate and (b) into the source as the drain resistance $R_D$ is varied, where $R_F = 5\:\mathrm{k}\Omega$ and the transistor's small-signal parameters are $g_m = 9.36\:\mathrm{mS}$ and $r_o = 14.4\:\mathrm{k}\Omega$. For (a), the source was loaded with $R_S = 100\:\Omega$; for (b), the gate was loaded with $R_G = 100\:\Omega$.}
\label{fig:RB-RE-Comp}
\end{figure*}

Fig.~\ref{fig:RB-RE-Comp}(b) shows the Spectre AC-simulated resistance looking into the source of a 36-nm \emph{n}-channel FinFET, plotted alongside the expression from Table~\ref{tab:MOS}'s fifth row.\footnote{The body effect is neglected here.} The asymptotic behaviors for $R_D = 0$ and $R_D \rightarrow\infty$, assuming $g_m r_o \gg 1$, are also indicated.

\subsection[“Open Collector/Drain” Behavior]{``Open Collector/Drain'' Behavior\footnote{The circuits under analysis here do not feature a ``true'' open collector or open drain---those terminals are ``open'' merely in a small-signal sense.}}
\label{sub:infRC}
It is instructive to compare Figs.~\ref{fig:RB-RE-Comp}(a) and (b) as $R_D\rightarrow\infty$. Notice how both $r_g - R_S$ and $r_s - R_G$ approach the same limit. This makes physical sense: if the drain is opened, the transistor (in the absence of the body effect\footnote{This caveat is quite important and worth pondering. In the body effect's presence, the equivalent gate-source impedance is ill-defined, since the circuit that attempts to define it leaves the (ac) ground terminal---which the body is tied to---floating. (More precisely, the circuit's nodal equations will yield an infinite number of solutions for $v_g - v_s$.) Indeed, without setting $g_{mb}=0$, the expressions for $r_g\!\left(R_D\rightarrow\infty\right)$ and $r_s\!\left(R_D\rightarrow\infty\right)$ would not coincide.\label{foot:infRD-nobody}}) can be viewed as a two-terminal device between the gate and the source whose equivalent small-signal impedance is given by
\begin{equation}
\begin{aligned}
r_{gs}\!\left(R_D\rightarrow\infty,g_{mb}=0\right) &\equiv r_g\!\left(R_D\rightarrow\infty,g_{mb}=0\right) - R_S \\ &\equiv r_s\!\left(R_D\rightarrow\infty,g_{mb}=0\right) - R_G \\[0.5em]
&= \frac{R_F+r_o}{1+g_mr_o}.
\end{aligned}
\label{eq:rgs-infRD}
\end{equation}
Analogous behavior is observed for the bipolar transistor. Referring to the last rows of Tables~\ref{tab:RB} and \ref{tab:RE}, we can write down the equivalent base-emitter impedance when $R_C\rightarrow\infty$:
\begin{equation}
\begin{aligned}
r_{be}\!\left(R_C\rightarrow\infty\right) &\equiv r_b\!\left(R_C\rightarrow\infty\right) - R_E \\ &\equiv r_e\!\left(R_C\rightarrow\infty\right) - R_B \\[0.5em]
&= r_{\pi} \biggm\Vert \frac{R_F+r_o}{1+g_mr_o} \\
&= \alpha r_m \biggm\Vert \frac{R_F+r_o}{1-g_mR_F}.
\end{aligned}
\label{eq:rbe-infRC}
\end{equation}
(The equivalence between $r_b$ and $r_e$ when the collector is opened is also noted in Tables~\ref{tab:RB} and \ref{tab:RE}.) Interestingly, this resistance is equal to its own base-emitter reciprocal: $r_{be}\!\left(R_C\rightarrow\infty\right) \rightleftharpoons r_{be}\!\left(R_C\rightarrow\infty\right)$.

\subsection[Base-Emitter Reciprocity]{\texorpdfstring{Base-Emitter Reciprocity Between $r_b$ and $r_e$}{Base-Emitter Reciprocity}}
As alluded to earlier in Section~\ref{subsec:cb-recip}, base-emitter reciprocity also appears in the impedance context. It is unsurprising that the $Y$-matrices on the left-hand sides of \eqref{eq:RB-nodal} and \eqref{eq:RE-nodal} are base-emitter reciprocals of each other [i.e., are related via \eqref{eq:BE-recip}]. One can easily see that the impedances looking into the base and into the emitter satisfy the following:
\begin{equation}
\begin{alignedat}{2}
&r_b \quad &&\rightleftharpoons \quad r_e \\
&r_b\left(R_E = 0\right) \quad &&\rightleftharpoons \quad r_e\left(R_B = 0\right) \\
&r_b\left(R_F\rightarrow\infty\right) \quad &&\rightleftharpoons \quad r_e\left(r_o\rightarrow\infty\right) \\
&r_b\left(r_o\rightarrow\infty\right) \quad &&\rightleftharpoons \quad r_e\left(R_F\rightarrow\infty\right) \\
&r_b\left(R_C\rightarrow\infty\right) \quad &&\rightleftharpoons \quad r_e\left(R_C\rightarrow\infty\right).
\end{alignedat}
\end{equation}

\begin{table*}[ht]
\centering
\caption{Resistance Looking into Emitter}
\label{tab:RE}
    \begin{NiceTabular}[width=\linewidth]{X[9,c,m] X[13,c,m] X[38,c]}[hvlines]
    General Expression & \vspace{2pt} \includegraphics[scale=0.4]{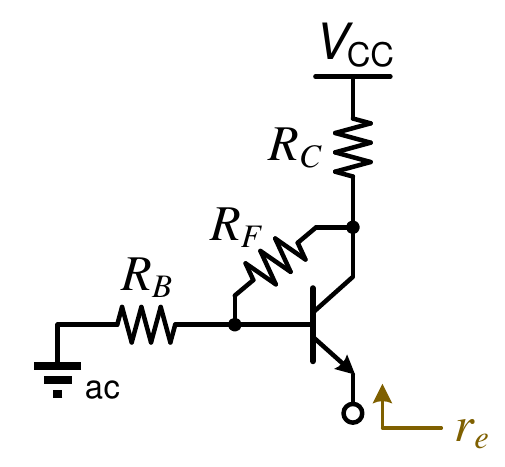} & \(\displaystyle r_e = \dfrac{r_o + \left[R_C\parallel (R_B+R_F)\right]+ g_m\left(\dfrac{R_F}{\beta} + \dfrac{R_C \parallel r_o}{\alpha}\right) \dfrac{R_B(R_C+r_o)}{R_B+R_C+R_F}}{1 + \dfrac{g_mr_o}{\alpha} + \dfrac{g_m\left[(R_B+R_C) \parallel R_F\right]}{\beta}} \) \\
    No Base Degeneration \linebreak $\left(R_B=0\right)$ & \vspace{2pt} \includegraphics[scale=0.4]{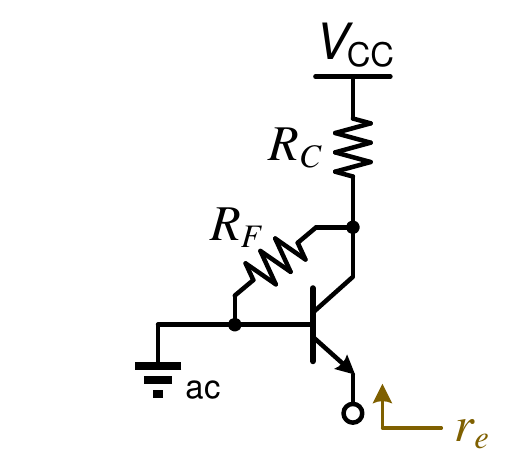} & \(\displaystyle r_e = \alpha r_m \biggm\Vert \dfrac{r_o + \left(R_C \parallel R_F\right)}{1 - g_m \left(R_C \parallel R_F\right)} = r_{\pi} \biggm\Vert \dfrac{r_o + \left(R_C \parallel R_F\right)}{1 + g_mr_o} \) \\
    \RowStyle[cell-space-limits=6pt]{} No Feedback \linebreak $\left(R_F \rightarrow \infty\right)$ & \includegraphics[scale=0.4]{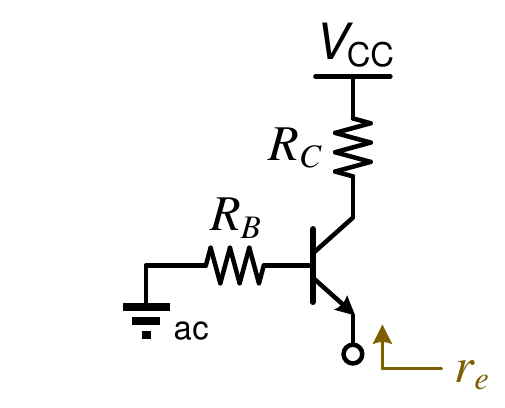} & \(\displaystyle \begin{aligned} r_e = \left(\alpha r_m + \dfrac{R_B}{\beta+1}\right) &\left\Vert \left( \dfrac{R_C+r_o}{1 - \dfrac{g_mR_C}{1+g_mR_B/\beta}}\right)\right. \\ = \left(r_{\pi} + R_B\right) &\left\Vert \left( \dfrac{R_C+r_o}{1 + \dfrac{g_mr_o}{1+g_mR_B/\beta}}\right)\right. \end{aligned} \) \\
    Neglecting Output Resistance \linebreak $\left(r_o \rightarrow \infty\right)$ & \vspace{2pt} \includegraphics[scale=0.4]{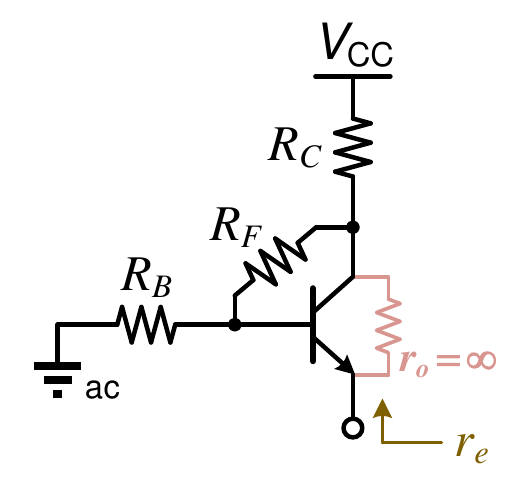} & \(\displaystyle r_e = \alpha r_m + \dfrac{R_B \left(R_C + \dfrac{R_F}{\beta+1}\right)}{R_B+R_C+R_F} \) \\
    Current-Source Bias at Collector \linebreak $\left(R_C\rightarrow\infty\right)$ & \vspace{2pt} \includegraphics[scale=0.4]{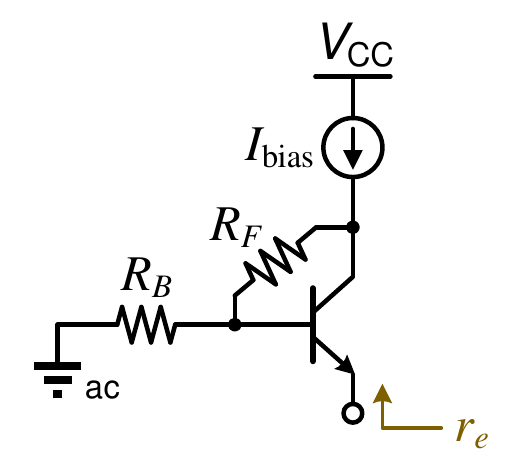} & \(\displaystyle r_e = \left(r_{\pi} \biggm\Vert \dfrac{R_F+r_o}{1+g_mr_o}\right) + R_B = \left(\alpha r_m \biggm\Vert \dfrac{R_F+r_o}{1-g_mR_F}\right) + R_B \) \linebreak\linebreak\linebreak (Compare with $r_b$ assuming $R_C\rightarrow\infty$.) \\
    \end{NiceTabular}
\end{table*}

\section{Impedance Looking into Collector}
\label{sec:rc}
Consider the circuit shown in the first row of Table~\ref{tab:RC}, wherein we are interested in obtaining the resistance $r_c$ looking into the collector; to do so, we apply a test current $i_T$ at the collector and probe the voltage across it, leading to the following nodal equations:
\begin{equation}
\begin{aligned}
\begin{pmatrix}
\dfrac{1}{R_F} + \dfrac{1}{r_o} & g_m - \dfrac{1}{R_F} & -g_m - \dfrac{1}{r_o} \\[1em]
-\dfrac{1}{R_F} & \dfrac{1}{r_{\pi}} + \dfrac{1}{R_B} + \dfrac{1}{R_F} & -\dfrac{1}{r_{\pi}} \\[1em]
-\dfrac{1}{r_o} & -g_m - \dfrac{1}{r_{\pi}} & g_m + \dfrac{1}{r_{\pi}} + \dfrac{1}{R_E} + \dfrac{1}{r_o}
\end{pmatrix}
\\[1em]
\begin{pmatrix}
v_c \\[0.5em] v_b \\[0.5em] v_e
\end{pmatrix}
=
\begin{pmatrix}
1 \\[0.5em] 0 \\[0.5em] 0
\end{pmatrix}
i_T.
\end{aligned}
\label{eq:RC-nodal}
\end{equation}
Solving for $r_c \coloneqq v_c/i_T$ yields the expression in the first row of Table~\ref{tab:RC}. Various approximations of interest are provided in the subsequent rows, and the analogous impedance looking into the drain of a MOSFET is given in the last row of Table~\ref{tab:MOS}.

A salient approximation of interest for $r_c$ pertains to the case with no feedback and no base degeneration; this situation allows designers to focus on how to use emitter degeneration to ``boost'' the output resistance of a transistor beyond $r_o$. With $R_B=0$ and $R_F\rightarrow\infty$, we have
\begin{equation*}
\begin{aligned}
\lim_{\substack{R_B = 0 \\ R_F\rightarrow\infty}} r_c &= \cfrac{r_o + R_E + \cfrac{g_mr_oR_E}{\alpha}}{1 + \cfrac{g_mR_E}{\beta}} \\[1em]
&= r_o + \left(1 + g_mr_o\right) \left(R_E\parallel r_{\pi}\right) \\[1em]
&= \begin{cases}
r_o + \left(1 + g_mr_o\right)R_E, &\beta\rightarrow\infty \\
r_{\pi} + \left(\beta+1\right)r_o, &R_E\rightarrow\infty
\end{cases}
\\[1em]
&\approx \begin{cases}
g_mr_oR_E, &\beta\rightarrow\infty \\
\beta r_o, &R_E\rightarrow\infty
\end{cases}
\end{aligned}
\end{equation*}
where we assumed $g_mr_o, \, g_mR_E, \, \beta \gg 1$ in the very last step. These approximations give rise to the conventional design wisdom associated with cascodes: stacking a MOS transistor above a current source enhances the current source's impedance by $g_mr_o$, and degenerating a bipolar transistor with a large impedance at its emitter enhances the transistor's output impedance by $\beta$.

Also, there is a simple, intuitive explanation for the resistance looking into the drain (Table~\ref{tab:MOS}'s last row). Current can flow through one of two paths here: through the series combination of $R_F$ and $R_G$, or through the transistor. Consider the latter. In the absence of the feedback resistor (i.e., $R_F\rightarrow\infty$), the impedance looking into the drain of a source-degenerated MOSFET is simply $r_o + R_S + \left(g_m + g_{mb}\right)r_oR_S$. If one shorts the drain and gate---``diode-connecting'' the transistor---this impedance is reduced by a factor of $1+g_mr_o$. Here, the diode connection is dulled by the voltage divider between $R_G$ and $R_F$: a voltage at the drain is reduced by a factor of $R_G/\left(R_G+R_F\right)$ at the gate. Putting these observations together leads to the expression for $r_d$.

\begin{table*}[ht]
\centering
\caption{Resistance Looking into Collector}
\label{tab:RC}
    \begin{NiceTabular}[width=\linewidth]{X[9,c,m] X[13,c,m] X[38,c]}[hvlines]
    General Expression & \includegraphics[scale=0.4]{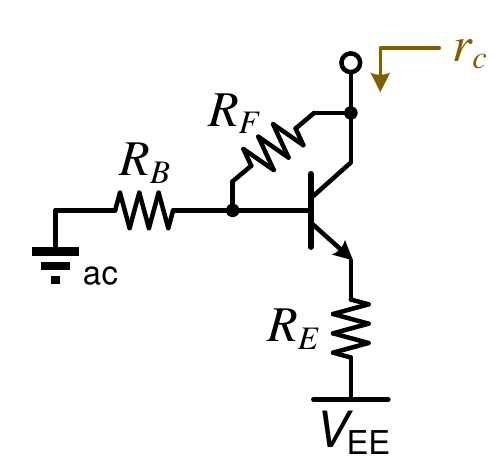} & \(\displaystyle r_c = \dfrac{\left(r_o + R_E + \dfrac{g_mr_oR_E}{\alpha}\right) \left(\dfrac{R_B+R_F}{r_o+R_F}\right) + \dfrac{g_mR_BR_F}{\beta}\left(\dfrac{r_o+R_E}{r_o+R_F}\right)}{1 + \left(1 + \dfrac{g_mr_o}{\alpha} + \dfrac{g_mR_F}{\beta}\right) \left(\dfrac{R_B+R_E}{r_o+R_F}\right)} \) \\
    No Base Degeneration \linebreak $\left(R_B=0\right)$ & \includegraphics[scale=0.4]{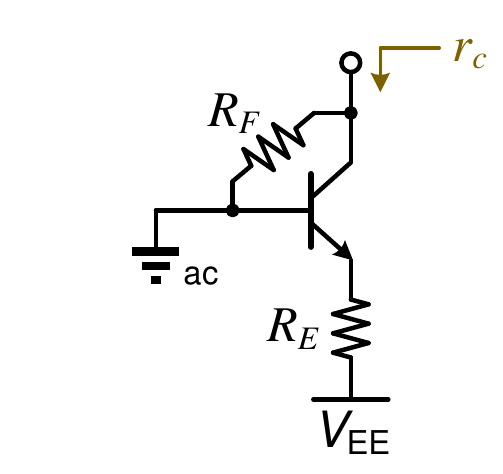} & \(\displaystyle r_c = R_F \left\Vert \left( \dfrac{r_o + R_E + \dfrac{g_mr_oR_E}{\alpha}}{1 + \dfrac{g_mR_E}{\beta}}\right) \right. \) \\
    No Emitter Degeneration \linebreak $\left(R_E=0\right)$ & \includegraphics[scale=0.4]{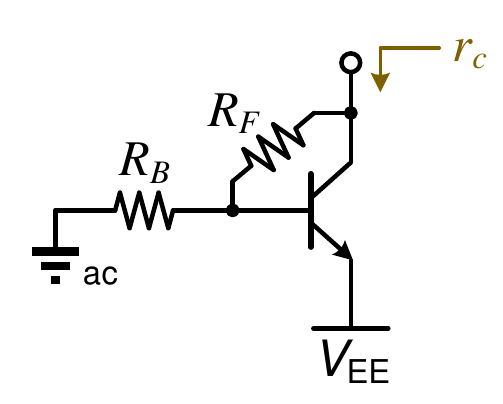} & \(\displaystyle r_c = r_o \left\Vert \left( \dfrac{R_B + R_F + \dfrac{g_mR_BR_F}{\beta}}{1 + \dfrac{g_mR_B}{\alpha}}\right) \right. \) \\
    No Feedback \linebreak $\left(R_F \rightarrow \infty\right)$ & \includegraphics[scale=0.4]{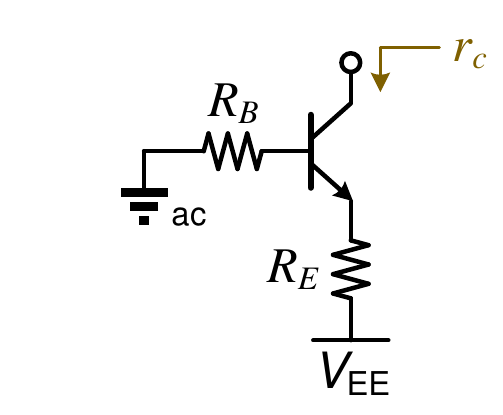} & \(\displaystyle r_c = \dfrac{\left(r_o + R_E\right) \left(1+\dfrac{g_mR_B}{\beta}\right) + \dfrac{g_mr_oR_E}{\alpha}}{1 + \dfrac{g_m(R_B+R_E)}{\beta}} \) \\
    Neglecting Output Resistance \linebreak $\left(r_o \rightarrow \infty\right)$ & \includegraphics[scale=0.4]{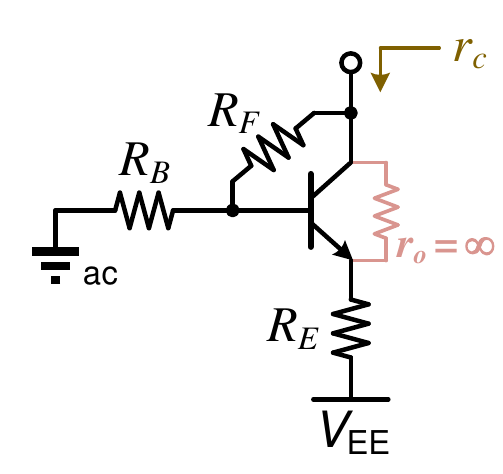} & \(\displaystyle r_c = \dfrac{\left(R_B + R_F\right) \left(1+\dfrac{g_mR_E}{\alpha}\right) + \dfrac{g_mR_BR_F}{\beta}}{1 + \dfrac{g_m(R_B+R_E)}{\alpha}} \) \\
    Current-Source Bias at Emitter \linebreak $\left(R_E\rightarrow\infty\right)$ & \includegraphics[scale=0.4]{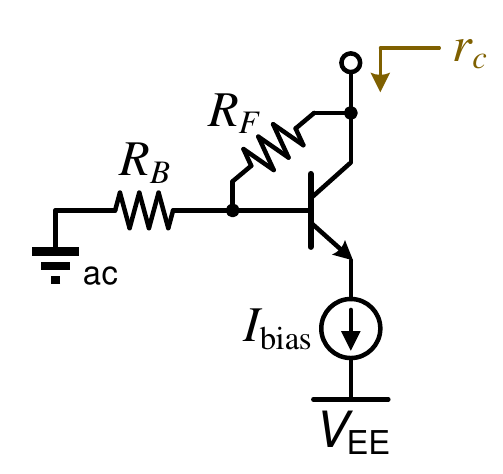} & \(\displaystyle r_c = \left[\left(r_{\pi} + \left(\beta+1\right)r_o\right) \!\parallel\! R_F\right] + R_B \) \linebreak\linebreak (Compare with $r_b$ assuming $R_E\rightarrow\infty$.) \\
    \end{NiceTabular}
\end{table*}

\begin{figure}[h]
\centering
\includegraphics[width=\columnwidth]{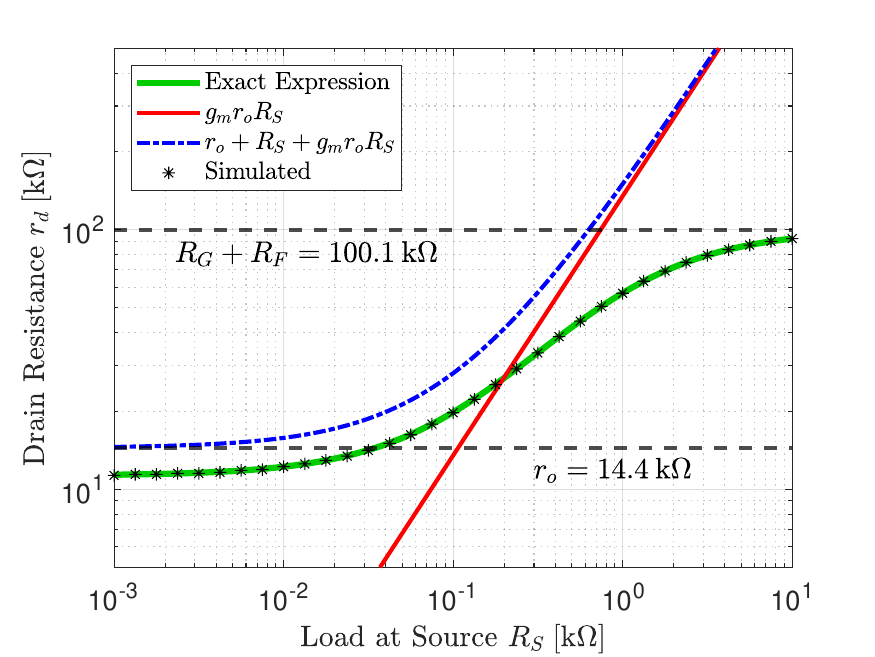}
\caption{Small-signal resistance looking into the drain as the source resistance $R_S$ is varied, where $R_G = 100\:\Omega$, $R_F = 100\:\mathrm{k}\Omega$, and the transistor's small-signal parameters are $g_m = 9.36\:\mathrm{mS}$ and $r_o = 14.4\:\mathrm{k}\Omega$.}
\label{fig:RC-Comp}
\end{figure}

Fig.~\ref{fig:RC-Comp} shows the Spectre AC-simulated resistance looking into the drain of a 36-nm \emph{n}-channel FinFET, plotted alongside the expression from Table~\ref{tab:MOS}'s last row.\footnote{The body effect is neglected here.} A couple familiar approximations as well as the asymptotic behaviors for $R_S = 0$ and $R_S \rightarrow \infty$ are depicted.

\subsection[“Open Emitter/Source” Behavior]{``Open Emitter/Source'' Behavior}
Finally, akin to our open collector/drain discussion in Section~\ref{sub:infRC}, we will briefly go over the open emitter/source scenarios. If $R_E \rightarrow \infty$, the following equivalent base-collector impedance is observed:
\begin{equation}
\begin{aligned}
r_{bc}\!\left(R_E\rightarrow\infty\right) &\equiv r_b\!\left(R_E\rightarrow\infty\right) - R_C \\ &\equiv r_c\!\left(R_E\rightarrow\infty\right) - R_B \\[0.5em]
&= \left[r_{\pi} + \left(\beta+1\right)r_o\right] \!\parallel\! R_F.
\end{aligned}
\end{equation}
(The equivalence between $r_b$ and $r_c$ when the emitter is opened is also noted in Tables~\ref{tab:RB} and \ref{tab:RC}.)

On the other hand, if $R_S \rightarrow \infty$, the only path for current to flow between the gate and drain terminals is not through the transistor at all, but through the external feedback resistor\footnote{In principal, we should also ignore the body effect here for this analysis to make sense (as we explained in footnote~\ref{foot:infRD-nobody}). But, it ultimately does not matter since the transistor vanishes from the picture entirely---no current can flow into \emph{any} of its terminals.} $R_F$:
\begin{equation}
\begin{aligned}
r_{gd}\!\left(R_S\rightarrow\infty\right) &\equiv r_g\!\left(R_S\rightarrow\infty\right) - R_D \\ &\equiv r_d\!\left(R_S\rightarrow\infty\right) - R_G \\[0.5em]
&= R_F.
\end{aligned}
\end{equation}

\subsection[Base-Emitter Reciprocity]{\texorpdfstring{Base-Emitter Reciprocity Within $r_c$}{Base-Emitter Reciprocity}}
Our final discussion of base-emitter reciprocity focuses not on comparing $r_c$ with some other impedance, but on $r_c$ itself. Notice how the expression for $r_c$ (Table~\ref{tab:RC}'s first row) is invariant under the swapping transformations of \eqref{eq:BE-recip}. The reason for this should be apparent from our discussion in Section~\ref{subsec:cb-recip}. Having the base and emitter terminals trade positions---which is what the set of transformations in \eqref{eq:BE-recip} does---has no impact on the $r_c$-measuring circuit under consideration here. It remains topologically unaltered.\footnote{Note that the swaps of \eqref{eq:BE-recip} permute \emph{both} the second and third rows \emph{and} the second and third columns of the admittance matrix in \eqref{eq:RC-nodal}---an operation that corresponds to permuting the second and third elements of the node-voltage vector, $v_b$ and $v_e$. Of course, this does not change the nodal equations' \emph{solution} for the first element, $v_c$.} Therefore, $r_c$ is its own base-emitter reciprocal, and we have the following base-emitter reciprocity relations:
\begin{equation}
\begin{alignedat}{2}
&r_c \quad &&\rightleftharpoons \quad r_c \\
&r_c\left(R_B = 0\right) \quad &&\rightleftharpoons \quad r_c\left(R_E = 0\right) \\
&r_c\left(R_F\rightarrow\infty\right) \quad &&\rightleftharpoons \quad r_c\left(r_o\rightarrow\infty\right).
\end{alignedat}
\end{equation}
It is worth noting that, within this context, the two transformations $g_m \Longleftrightarrow -g_m$ and $\alpha \Longleftrightarrow -\beta$ can be collapsed into a single transformation: $\alpha \Longleftrightarrow \beta$. This is because $g_m$ only ever appears in the expression for $r_c$ when divided by $\alpha$ or by $\beta$.

\section{Conclusion}
Expressions for the small-signal, low-frequency voltage gains of the common emitter/source, common base/gate, and common collector/drain voltage gains under the most general set of (resistive) loads were given. Expressions for the small-signal, low-frequency input and output impedances of these amplifiers were also provided. A unifying view of each topology that can simultaneously describe the effects of different design choices (such as emitter degeneration or feedback) was therefore presented. This bank of expressions should prove useful to both students learning analog electronics for the first time as well as circuit-design engineers wishing to bolster their intuition with a touch of mathematical rigor. Excellent agreement with simulation results is observed. 

We saw that, from a small-signal standpoint, exchanging the base and emitter terminals is mathematically equivalent to (1)~reversing the direction of the $g_m$ dependent current source, (2) swapping $\alpha$ and $-\beta$, and (3) swapping $r_o$ with $R_F$. This insight, which we termed ``base-emitter reciprocity,'' allows us to convert between the common emitter and common base gains, as well as between the impedances looking into the base and into the emitter, without analyzing the circuit anew. 

Future work might focus on further generalizing the collection of results assembled herein to capture frequency-dependent behavior, as well as seeing if a high-frequency extension of base-emitter reciprocity can be formulated. While we could account for the base-emitter and base-collector capacitances, $C_{\pi}$ and $C_{\mu}$, by respectively replacing $r_{\pi}$ with $r_{\pi} \parallel \left(1/sC_{\pi}\right)$ and $Z_F$ (now the feedback \emph{impedance}) with $Z_F \parallel \left(1/sC_{\mu}\right)$, a more elegant approach might involve incorporating the concept of time and transfer constants into the analysis \cite{TTC:OG,TTC}. Also of value could be an interpretation of this paper's analytical conclusions that foregrounds feedback-based viewpoints \cite{Blackman,EET}.\footnote{As a simple example, we can use Blackman's theorem \cite{Blackman} to calculate $r_b\!\left(R_E=0\right)$: with the $g_m$ source disabled, the impedance of interest becomes $Z_0 = r_{\pi} \parallel \left[R_F + \left(R_C \parallel r_o\right)\right]$; whereas the short- and open-circuit return ratios are $T_{\mathrm{sc}}=0$ and $T_{\mathrm{oc}} = g_m\left[\left(r_{\pi} + R_F\right) \parallel R_C \parallel r_o\right] \cdot \left[r_{\pi} / \left(r_{\pi} + R_F\right)\right]$, respectively. Then $r_b\!\left(R_E=0\right) = Z_0 \cdot \left(1 + T_{\mathrm{sc}}\right) / \left(1 + T_{\mathrm{oc}}\right)$, which can easily be simplified to the expression found in the second row of Table~\ref{tab:RB}.}

\begin{table*}[ht]
\centering
\caption{Expressions for MOS Transistors}
\label{tab:MOS}
    \begin{NiceTabular}[width=\linewidth]{X[11,c,m] X[13,c,m] X[36,c]}[hvlines]
    Common Source Amplifier & \includegraphics[scale=0.4]{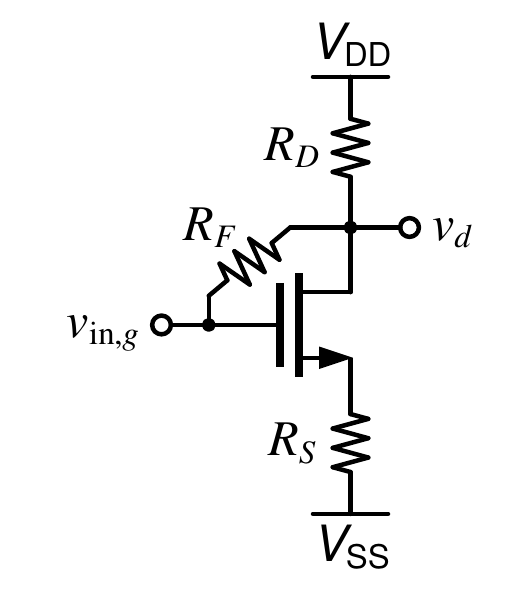} \vspace{-2pt} & \(\displaystyle A_v^{\mathrm{CS}} = -\dfrac{g_m \left[\dfrac{1}{1+\left(g_m+g_{mb}\right)R_S+R_S/r_o}\right]-\dfrac{1}{R_F}}{\dfrac{1}{R_D \parallel R_F} + \dfrac{1}{r_o} \left[\dfrac{1}{1+\left(g_m+g_{mb}\right)R_S + R_S/r_o}\right]} \) \\
    Common Gate Amplifier & \includegraphics[scale=0.4]{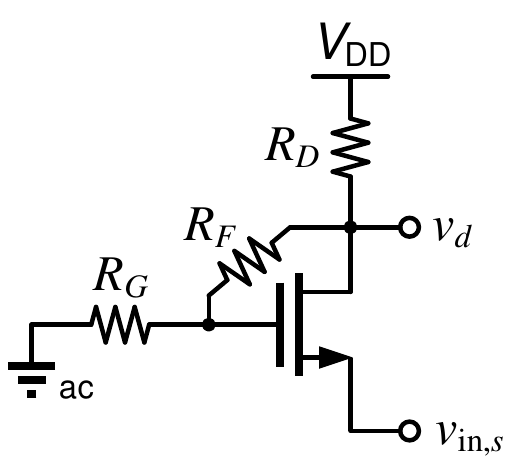} & \(\displaystyle A_v^{\mathrm{CG}} = \dfrac{g_m + g_{mb} + \dfrac{1}{r_o}}{\dfrac{1}{R_D \parallel r_o} + \dfrac{1+g_mR_G}{R_G+R_F}} \) \\
    Common Drain Amplifier \linebreak (Source Follower) & \includegraphics[scale=0.4]{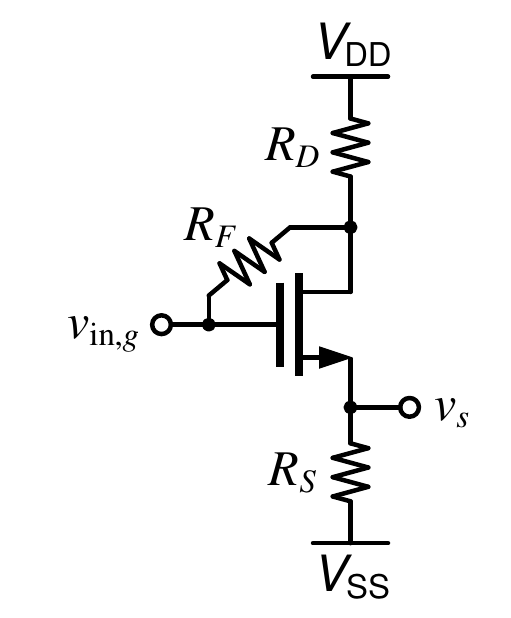} \vspace{-2pt} & \(\displaystyle A_v^{\mathrm{CD}} = \dfrac{g_mr_o + \dfrac{R_D}{R_D+R_F}}{\left(g_m + g_{mb}\right)r_o + \dfrac{r_o + \left(R_D \parallel R_F\right)}{R_S} + 1} \) \\
    Resistance Looking into Gate & \includegraphics[scale=0.4]{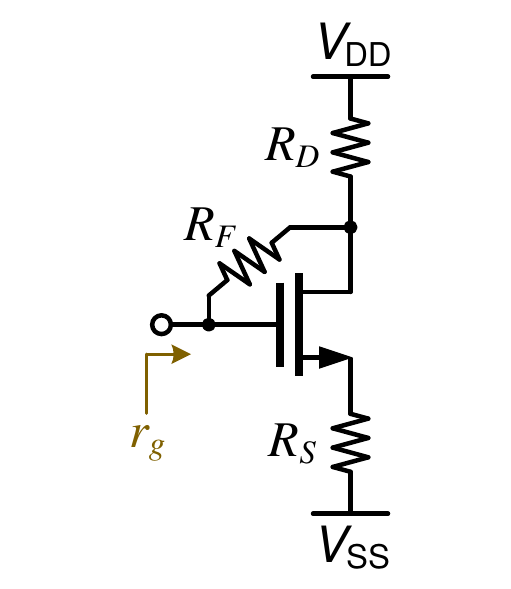} \vspace{-2pt} & \(\displaystyle r_g = R_F + \left[ R_D \left\Vert \dfrac{r_o + R_S + g_mr_oR_S \left(1 + \dfrac{g_{mb}}{g_m} - \dfrac{R_F}{R_S}\right)}{1 + g_mr_o \left(1 + \dfrac{R_F}{R_D}\right)} \right. \right] \) \\
    Resistance Looking into Source & \includegraphics[scale=0.4]{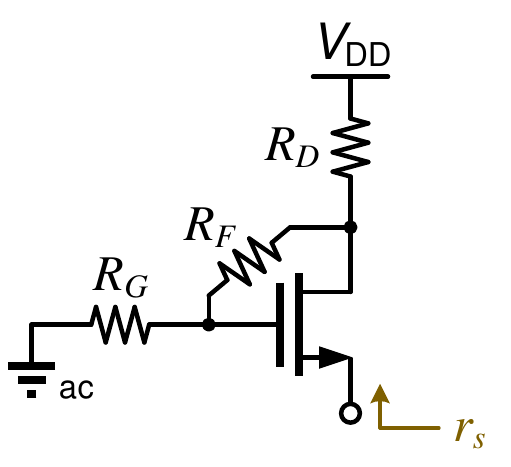} & \(\displaystyle r_s = \left( r_m \parallel r_{mb} \parallel r_o\right) + \dfrac{R_D}{R_G + R_D + R_F} \left[\dfrac{\left(1+g_mr_o\right)R_G + R_F}{1 + \left(g_m+g_{mb}\right)r_o}\right] \) \\
    Resistance Looking into Drain & \includegraphics[scale=0.4]{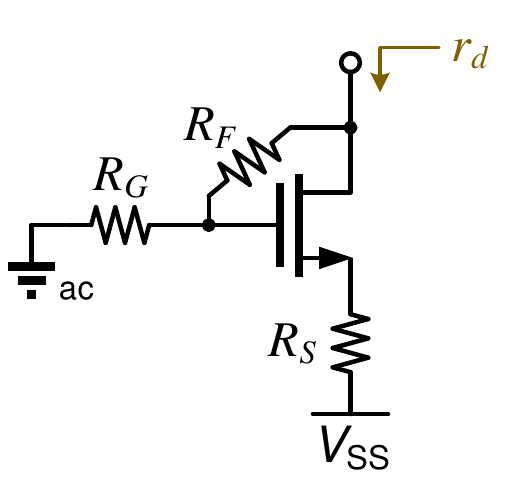} & \(\displaystyle r_d = \left(R_G+R_F\right) \left\Vert \left[\dfrac{r_o + R_S + \left(g_m + g_{mb}\right)r_oR_S}{1 + g_mr_o\left(\dfrac{R_G}{R_G+R_F}\right)}\right]\right. \) \\
    \end{NiceTabular}
\end{table*}


\ifCLASSOPTIONcaptionsoff
  \newpage
\fi

\end{document}